\documentclass{JHEP3}   
\usepackage{epsfig}

\newcommand{\fref}[1]{Figure~\ref{#1}}
\newcommand{\cref}[1]{Chapter~\ref{#1}}
\newcommand{\beq}{\begin{equation}}
\newcommand{\eeq}{\end{equation}}
\newcommand{\ba}{\begin{array}}
\newcommand{\ea}{\end{array}}
\newcommand{\bcenter}{\begin{center}}
\newcommand{\ecenter}{\end{center}}

\def\IB{\relax\hbox{$\inbar\kern-.3em{\rm B}$}}
\def\IC{\relax\hbox{$\inbar\kern-.3em{\rm C}$}}
\def\ID{\relax\hbox{$\inbar\kern-.3em{\rm D}$}}
\def\IE{\relax\hbox{$\inbar\kern-.3em{\rm E}$}}
\def\IF{\relax\hbox{$\inbar\kern-.3em{\rm F}$}}
\def\IG{\relax\hbox{$\inbar\kern-.3em{\rm G}$}}
\def\IGa{\relax\hbox{${\rm I}\kern-.18em\Gamma$}}
\def\IH{\relax{\rm I\kern-.18em H}}
\def\IK{\relax{\rm I\kern-.18em K}}
\def\IL{\relax{\rm I\kern-.18em L}}
\def\IP{\relax{\rm I\kern-.18em P}}
\def\IR{\relax{\rm I\kern-.18em R}}
\def\IZ{\relax\ifmmode\mathchoice
{\hbox{\cmss Z\kern-.4em Z}}{\hbox{\cmss Z\kern-.4em Z}}
{\lower.9pt\hbox{\cmsss Z\kern-.4em Z}}
{\lower1.2pt\hbox{\cmsss Z\kern-.4em Z}}\else{\cmss Z\kern-.4em Z}\fi}
\def\II{\relax{\rm I\kern-.18em I}}

\def\sCC{{\kern 0.27em\vrule height1.45ex width0.03em depth0em
          \kern-0.30em\rm C}}
\def\C{{\mathchoice
  {\sCC}
  {\sCC}
  {\kern 0.225em \vrule height1.05ex width0.025em depth0em \kern-0.25em \rm C}
  {\kern 0.180em \vrule height0.78ex width0.02em depth0em \kern-0.2em \rm C}
        }}
\def\sHH{{\rm I\kern-.16em{}H}}
\def\H{{\mathchoice
  {\sHH}
  {\sHH}
  {\rm I\kern-.13em{}H}
  {\rm I\kern-.13em{}H} }}
\def\sNN{{\rm I\kern-.16em{}N}}
\def\N{{\mathchoice
  {\sNN}
  {\sNN}
  {\rm I\kern-.12em{}N}
  {\rm I\kern-.10em{}N} }}
\def\sPP{{\rm I\kern-.16em{}P}}
\def\P{{\mathchoice
  {\sPP}
  {\sPP}
  {\rm I\kern-.12em{}P}
  {\rm I\kern-.10em{}P} }}
\def\sQQ{{\kern 0.27em \vrule height1.45ex width0.03em depth0em
          \kern-0.30em \rm Q}}
\def\Q{{\mathchoice
        {\sQQ}
        {\sQQ}
  {\kern 0.225em \vrule height1.05ex width0.025em depth0em \kern-0.25em \rm Q}
  {\kern 0.180em \vrule height0.78ex width0.020em depth0em \kern-0.20em \rm Q}
        }}
\def\sRR{{\rm I\kern-0.16em{}R}}
\def\R{{\mathchoice
  {\sRR}
  {\sRR}
  {\rm I\kern-0.12em{}R}
  {\rm I\kern-0.10em{}R} }}
\def\sZZ{{\rm Z\kern-0.32em{}Z}}
\def\Z{{\mathchoice
  {\sZZ}
  {\sZZ} 
  {\rm Z\kern-0.3em{}Z}     %.3
  {\rm Z\kern-0.25em{}Z} }}  %.25
\def\ZZZ{{\rm Z\kern-0.24em{}Z}}
\def\sII{{\rm I\kern-0.16em{}I}}
\def\I{{\mathchoice
  {\sII}
  {\sII}
  {\rm I\kern-0.12em{}I}
  {\rm I\kern-0.10em{}I} }}

\def\inbar{\,\vrule height1.5ex width.4pt depth0pt}
\font\cmss=cmss10 \font\cmsss=cmss10 at 7pt

\def\smiley{\hbox{\large$\bigcirc$\hspace{-0.80em}\raise.2ex
\hbox{$\cdot\cdot$}\kern-.61em\lower.2ex\hbox{\scriptsize$\smile$}}\ }
\def\frowny{\hbox{\large$\bigcirc$\hspace{-0.80em}\raise.2ex
\hbox{$\cdot\cdot$}\kern-.635em\lower.2ex\hbox{\scriptsize$\frown$}}\ }

\def\I{{\rlap{1} \hskip 1.6pt \hbox{1}}}

\makeatletter
\let\hangafter\@hangfrom
\makeatother

\newcommand{\be}{\begin{equation}}
\newcommand{\ee}{\end{equation}}
\newcommand{\bea}{\begin{eqnarray}}
\newcommand{\eea}{\end{eqnarray}}
\newcommand{\bean}{\begin{eqnarray*}}
\newcommand{\eean}{\end{eqnarray*}}
\newcommand{\nn}{\nonumber}
\newcommand{\IN}{{\cal N}}

\preprint{MIT-CTP-3611\\
\baselineskip12pt\hbox{hep-th/0503177}}

\title{A New Infinite Class of Quiver Gauge Theories}

\author{Amihay Hanany, Pavlos Kazakopoulos, Brian Wecht\\

\vspace{0.3 cm}

{Center for Theoretical Physics,
                   Massachusetts Institute of Technology,\\
                   Cambridge, MA 02139, USA.}\\
~\\
~\\
\email{hanany, noablake, bwecht@mit.edu}
}

\abstract{We construct a new infinite family of ${\cal N}=1$
quiver gauge theories which 
can be Higgsed to the $Y^{p,q}$ quiver gauge theories. The
dual
geometries are toric Calabi-Yau cones for which we
give the toric data.
We also
discuss the action of Seiberg duality on these quivers, and 
explore the different Seiberg dual theories. We describe the relationship 
of these theories to five dimensional gauge theories on (p,q) 5-branes.
Using the toric data, we specify some of the properties of the
corresponding dual Sasaki-Einstein manifolds. These theories generically have
algebraic R-charges which are not quadratic irrational numbers.
The metrics for these manifolds still remain unknown.}

\keywords{Quiver Gauge Theories, Five Dimensional Gauge Theories}

\begin{document}

%%%%%%%%%%%%%%%%%%%%%%%%%%%%%%%%%%%%%%%%%%%%%%%%%%%%%%

%=====================================================
\section{Introduction}
%=====================================================
The past few months have brought exciting new developments in the study
of $\IN=1$ superconformal field theories. 
It is known that such field theories can be constructed in string theory
by placing a stack of D3-branes at the tip of a non-compact Calabi-Yau cone;
the first such examples of these theories were given for Calabi-Yau spaces which
are orbifolds of $\IC^3$ \cite{Douglas:1996sw}. More
generally,
the AdS/CFT correspondence \cite{maldacena,GKP,Witten:1998qj}
says that there is a gauge/string duality
which can be seen by
taking the near horizon limit of the D3-branes.
The simplest example, where the branes are
placed in flat ten dimensional space, gives a duality between
type IIB string theory on $AdS_5 \times S^5$ and $\IN=4$ SUSY Yang-Mills.
More generally, however, we can place the D3-branes at the tip
of a Calabi-Yau cone whose base is a five-dimensional 
Einstein-Sasaki manifold $X_5$; the near-horizon
limit of this is then Type IIB on $AdS_5 \times X_5$ \cite{Morrison:1998cs,Kehagias:1998gn}. 
This setup breaks additional supersymmetry,
generically leaving an $\IN=1$ theory on the D3-branes. These
$\IN=1$ theories are quiver gauge theories \cite{Douglas:1996sw}, i.e. theories
with product gauge groups and matter transforming in bifundamental 
representations.

Ideally, one would like to know the metric on the Calabi-Yau manifold, since this would
provide a great deal of information (e.g. volumes of calibrated submanifolds) 
which could be translated into
data about the field theory. In practice, however, it is a difficult task to find these
metrics, and until recently the metrics on only two 
Einstein-Sasaki five-manifolds were known: $S^5$ and $T^{1,1}$. Recently,
an infinite family of Einstein-Sasaki manifolds with topology $S^2 \times S^3$
were found by Gauntlett, Sparks, Martelli and Waldram \cite{Gauntlett:2004yd};
these extend
previous work by these authors on compactifications of M-theory \cite{Gauntlett:2004zh}.
The new Einstein-Sasaki spaces have been dubbed $Y^{p,q}$,
where $p$ and $q$ are integers with $0 < q < p$; the Calabi-Yau cones over
these spaces are toric.
Sparks and Martelli \cite{Martelli:2004wu} recently computed the corresponding
toric diagrams and found that $Y^{2,1}$ was in fact the horizon of the complex cone
over the first del Pezzo surface $dP_1$. The corresponding gauge theory is known,
having been computed via partial resolution of the orbifold $\IC^3/\IZ_3 \times \IZ_3$
\cite{Feng:2000mi}. This begged the question: What are the gauge theories
dual to the other $Y^{p,q}$?

The problem of finding these gauge theories was quickly solved. 
The authors of \cite{BFHMS}
constructed the dual $\IN=1$ gauge theories (also called $Y^{p,q}$), 
thus providing an infinite family of AdS/CFT duals. The $Y^{p,q}$ theories
have gauge group $SU(N)^{2p}$, and the complete matter content and
superpotentials are known.
These theories have survived
many nontrivial checks of the AdS/CFT duality, 
such as matching volumes of 3-manifolds in the Calabi-Yau with dimensions of 
operators in the gauge theory computed via $a$-maximization
\cite{Intriligator:2003jj}. The $a$-maximization calculation was done
for $Y^{2,1}$ in \cite{Bertolini:2004xf}, and the authors of
\cite{BFHMS} found a general formula for the R-charges for any $Y^{p,q}$; these
agree beautifully with the corresponding volume calculations in the string dual.  
For interesting related work on these $Y^{p,q}$ theories, 
see \cite{Franco:2005fd, Bergman:2005ba, 
Benvenuti:2005wi, Pal:2005mr, Benvenuti:2004wx, Franco:2004bq, Herzog:2004tr}.

Some of the $Y^{p,q}$ theories and geometries are already familiar. We can
formally extend the range of $q$ to $0 \leq q \leq p$ and find that $Y^{p,p} = S^5/ \IZ_{2p}$,
and $Y^{p,0} = T^{1,1}/\IZ_p$. These spaces are not smooth, but can be used as
interesting string backgrounds. The
corresponding gauge theories to these spaces are well-known
\cite{Douglas:1996sw,Feng:2000mi,Klebanov:1998hh,Feng:2001xr}, and the
rest of the $Y^{p,q}$ gauge theories were constructed by a simple procedure
that generalizes the method for going from $Y^{2,2}=S^5/\IZ_2\times\IZ_2$ 
to $Y^{2,1}=dP_1$ to $Y^{2,0}=\IF_0$.

The fact that the $Y^{p,q}$ spaces are toric is crucial to the above calculations.
The toric diagram for a given geometry
can be described as points in a three-dimensional lattice; these points describe a
$\C^*$ action and uniquely specify the resulting geometry. In this sense, toric geometries
are generalizations of $\C \P^n$. Equivalently, we can describe the geometry
with a {\it toric fan}, which is a series of vectors ending on the specified lattice points.
A toric Calabi-Yau cone then satisfies the additional restriction
that the endpoints of these vectors lie in a plane. For this reason, all the toric
diagrams we study can be represented on a two-dimensional lattice.  For
good reviews of toric geometry, see \cite{Leung:1997tw,fulton,Aharony:1997bh}. 

The construction of quiver gauge theories from toric singularities is a difficult task 
and many attempts at getting a general formula have been made so far with some or partial success. 
The general problem is to start with a geometric description, given by the toric data, and then to 
use it in order to find the corresponding gauge theory that lives on a D3 brane probing
the corresponding Calabi-Yau cone.
There are a few steps in finding the quiver gauge theory: First, one looks for the 
gauge groups. The second step is to determine the matter bifundamental fields, and finally 
one must find the corresponding superpotential which encodes the interaction terms. We
have described these three steps in order by level of difficulty;
the third step is often really a hard problem to solve.
If for some reason in an independent computation we have a good guess for the quiver 
gauge theory then it is a straightforward task to compute the corresponding toric data
\cite{Feng:2000mi}. 
However, the current known techniques are very long in computation time and are practical 
only for relatively small (few gauge groups and few matter fields) quiver gauge theories.

Until the computation of the $Y^{p,q}$ quiver gauge theories there were only 
finitely many toric singularities with known quiver gauge theories, most of 
which were centered around del Pezzo surfaces and toric diagrams with one or no 
internal points. This situation changed when the $Y^{p,q}$'s 
were found. This leads to the hope that there are many more such 
infinite families of quiver gauge theories for which the computation is a 
relatively simple task. This is going to be the point of this paper - 
to introduce another infinite family of quiver gauge theories for
which we know the toric data.

Shortly after the computation of the $Y^{p,q}$ quivers it was realized that 
the toric diagrams had actually been studied before in a seemingly unrelated environment. 
In \cite{Aharony:1997bh,Aharony:1997ju} the $Y^{p,q}$ toric diagrams were shown to be dual to a 
5-brane web which describes the dynamics of a five dimensional SYM $SU(p)$ gauge 
theory with 8 supercharges, with $q$ denoting the coefficient of a Chern Simons term. 
Only a few computations have been made since then 
\cite{Iqbal:2002ep,Iqbal:2003zz} and many more may be found. Using the 
connection to five dimensional gauge theories, we are able to borrow ideas 
from the construction of such theories using branes. In particular the 
process of adding flavors to the five dimensional gauge theory is a useful tool. 
As we will see this will be interpreted as an inverse Higgs mechanism (unhiggsing) 
for the quiver gauge theories and essentially is the main tool which allows 
the construction of the new infinite family of quiver theories. The correspondence 
with five dimensional theories thus turns out to be useful and may be 
used in future attempts to find other quiver theories.

In this paper, we describe a new infinite class of toric geometries and the
corresponding gauge theories. To motivate this construction, consider the
case of the (complex cone over the) second del Pezzo surface $dP_2$. Since
$dP_2$ is simply $\IP^2$ blown up at two points, it is easy to see that
one can blow down an exceptional $\IP^1$ to arrive at $dP_1$. This fact
is reflected in the corresponding gauge theories via Higgsing of
the $dP_2$ theory to the $dP_1$ theory by giving an appropriate bifundamental
field a vacuum expectation value. 
Additionally, however, one can also blow down $dP_2$ to $\IF_0$. As with $dP_1$,
this can also be seen from the gauge theory via Higgsing. Thus, $dP_2$ is an example
of a theory which can be Higgsed to give two different $Y^{p,q}$ theories. The goal
of the present paper is to generalize this construction. The analogous phenomenon in the 
context of five dimensional gauge theories with $SU(2)$ gauge group was studied in 
\cite{Aharony:1997bh,Aharony:1997ju,Douglas:1996xp}, where it is associated with a 
discrete $\theta$ angle and the limit of a gauge theory with one flavor that has a 
large mass. The sign of the mass then implies which of the resulting theories is either $\IF_0$ or $dP_1$.

We find a general procedure for constructing 
four dimensional ${\cal N}=1$ $SU(N)^{2p+1}$ gauge theories we denote as $X^{p,q}$, 
which Higgs to $Y^{p,q}$ and 
$Y^{p,q-1}$.  We present the toric diagrams, quivers, and superpotentials for these theories. 
The Higgsing process can be seen in a number of
ways; the toric diagrams, quiver, and (p,q) 5-brane webs all provide illuminating
perspectives, which we describe in detail. In addition, we discuss the different toric phases 
of the $X^{p,q}$ theories, i.e. the quivers one can get by Seiberg duality
that are still  $SU(N)^{2p+1}$ gauge theories. 

The outline of this paper is as follows. In Section 2, we motivate the construction
of the $X^{p,q}$ theories by doing two simple examples. We first explore the
relationship between $dP_2$, $dP_1$, and $\IF_0$ from
both the gauge theory and the toric perspectives. In addition,
we describe an analogous relationship between the Suspended Pinch Point (SPP),
conifold, and $\IC^3/\IZ_2$. In Section 3,
we describe the construction of one phase of the $X^{p,q}$ spaces. It 
is necessary to treat two cases, $q=p$ and $1 \leq q \leq p-1$, which we do.
As a detailed example, we write down the quivers and superpotentials for 
the $X^{3,1}$, $X^{3,2}$, and $X^{3,3}$ theories. Finally, we discuss the toric
diagrams for the dual $X^{p,q}$ geometries.
In Section 4,
we review the relationship of toric geometry with webs of (p,q) 5-branes, and 
describe how the Higgsing process can be seen from this perspective. We discuss
how many parameters of the four dimensional theory are related to parameters
of the five dimensional theory. 
In Section 5, we discuss the Seiberg dual phases of the $X^{p,q}$ theories that
still have gauge group $SU(N)^{2p+1}$; these are
usually called the ``toric phases'' of the theory.
There is a general relationship between
the number of bifundamentals in the different toric phases of the $X^{p,q}$ with
the number of bifundamentals in the different toric phases of $Y^{p,q}$ and 
$Y^{p,q-1}$, which we discuss. Finally, in section 6, we discuss the R-charges for
the $X^{p,q}$ theories. The calculation is in general computationally quite difficult but can be
done for some small values of $p$. We calculate the R-charges for $X^{2,2}$ and $X^{3,3}$ 
and find that they not quadratic irrational but instead the roots of
quartic equations. In addition, we discuss how although $X^{2,1}$ has
a two-dimensional basis of R-charges, this property does not appear
to be true for larger values of $p$. This remains puzzling in light of recent results
\cite{djy}.

%=====================================================
\section{Warm-Ups: del Pezzo 2 and the Suspended Pinch Point}
%=====================================================

Before we proceed to the general construction, we review the case of the gauge theory
dual to the cone over $dP_2$. This theory provides a template for
constructing the more general quivers, and as such is a useful example to explore.

\begin{figure}[ht]
  \epsfxsize = 6cm
  \centerline{\epsfbox{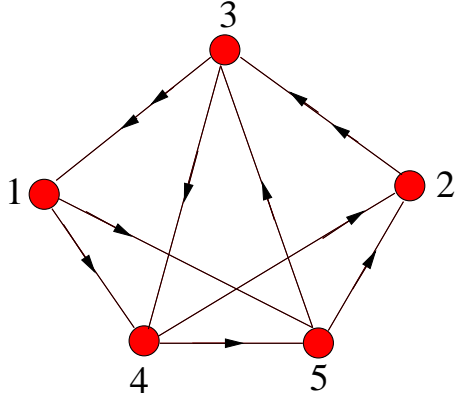}}
  \caption{One quiver for the $dP_2$ gauge theory.}
  \label{quiver_dp2}
\end{figure}

Although we do not know the explicit metric on $dP_2$, the gauge theory has been
constructed by partially resolving the space $\IC^3 / \IZ_3 \times \IZ_3$ 
\cite{Feng:2000mi,Feng:2001xr}. One quiver
for this theory is given in Figure \ref{quiver_dp2}. 

The accompanying superpotential is given by 
\bea
W & = & X_{34}X_{45}X_{53}-X_{53}Y_{31}X_{15}-X_{34}X_{42}Y_{23} \\
& + & Y_{23}X_{31}X_{15}X_{52}+X_{42}X_{23}Y_{31}X_{14}-X_{23}X_{31}X_{14}X_{45}X_{52}.
\label{wdp2}
\eea
This superpotential has 3 cubic terms, 2 quartics, and one quintic. 

\begin{figure}[ht]
  \epsfxsize = 10cm
  \centerline{\epsfbox{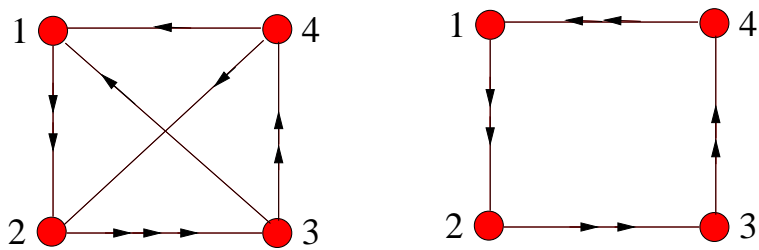}}
  \caption{The quivers for $Y^{2,1}$ (left) and $Y^{2,0}$ (right)}
  \label{quiver_y21y20}
\end{figure}

For our purposes in this work, the interesting thing to notice about the $dP_2$ quiver is that
it can be Higgsed to either $Y^{2,1}$ or $Y^{2,0}$ by giving a vev to 
(for example) $X_{52}$ or $X_{45}$, respectively. See Figure \ref{quiver_y21y20} for these quivers.
The Higgsing is straightforward: Giving a vev to a bifundamental field simply breaks the
$SU(N) \times SU(N)$ group under which it transforms to the diagonal. This means
that we should combine those two nodes in the quiver and delete the bifundamental
from the theory this flows to in the IR. If there is a cubic term with the
bifundamental in it, then we should also delete the other bifundamental
fields; the vev will give them a mass and they should be integrated out
of the IR theory.

To be complete, we must also check that the superpotential behaves as it
should under Higgsing. Giving a vev to $X_{52}$ is straightforward: Since 
this bifundamental only appears in terms with four or more fields, 
we simply delete it from the appropriate terms and relabel the remaining
fields. This yields the superpotential
\bea
W_{dP_1} & = & X_{31}Y_{12}X_{23}-X_{23}Y_{34}X_{42}-X_{31}X_{12}Y_{23} \\
&+& Y_{23}X_{34}X_{42}+X_{12}Z_{23}Y_{34}X_{41}-Z_{23}X_{34}X_{41}Y_{12}. 
\label{wdp1}
\eea
This is the superpotential for the gauge theory dual to the complex cone
over the first del Pezzo surface $dP_1$. Note that we have relabelled some
nodes for aesthetic reasons.

Now, let's see what happens when we set $\langle X_{45} \rangle  \neq 0$. Since
$X_{45}$ appears in a cubic term $X_{45}X_{53}X_{34}$ in the $dP_2$ superpotential, this
vev has the effect of giving a mass to $X_{53}$ and $X_{34}$. These fields
should then be integrated out of the IR theory. The classical equations of motion
are

\bea
{\partial W \over \partial X_{53}}  =  X_{45}-Y_{31}X_{15} = 0, \qquad
{\partial W \over \partial X_{34}}  =  X_{53}-X_{42}Y_{23} = 0.
\label{dp2eom}
\eea

The resulting superpotential is then purely quartic and is given by
\bea
W_{\IF_0} = X_{12}Y_{23}X_{34}Y_{41}-X_{12}X_{23}X_{34}X_{41}-Y_{12}Y_{23}Y_{34}Y_{41}
+Y_{12}X_{23}Y_{34}X_{41},
\eea
which is indeed the superpotential for $\IF_0$. 
Notice that $dP_2$ has 11 fields, $dP_1$ has 10, and $\IF_0$ has 8. The cubic term
has had the effect of removing two additional fields from the spectrum, as it must. 

The R-charges for this theory must be computed via $a$-maximization \cite{Intriligator:2003jj}. 
There are {\it a priori} 11 different R-charges, subject to 5 constraints from
anomaly freedom and 6 constraints from the superpotential. One can easily
check that there are, however, 4 undetermined R-charges; the maximization
must then be done over the space of these 4 charges \cite{Bertolini:2004xf}. This
will be a general feature of our new quivers. 

\begin{figure}[ht]
  \epsfxsize = 10cm
  \centerline{\epsfbox{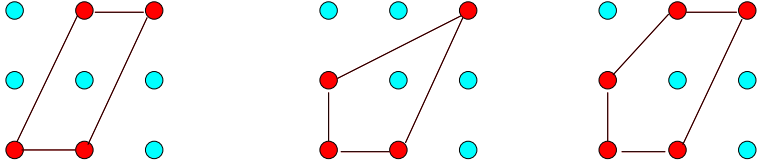}}
  \caption{The toric data for $\IF_0$ (left), $dP_1$ (center), and $dP_2$ (right).}
  \label{toric1}
\end{figure}

Let's now recall the toric presentations of the complex cones over $\IF_0, dP_1,$ and $dP_2$. 
Since these cones are Calabi-Yau, we may represent the toric data with a two
dimensional lattice. A 
perhaps familiar presentation of the toric data for these three surfaces is given in
Figure \ref{toric1}.  
The toric diagram can be used to read off the number of 
2- and 4-cycles in the corresponding Calabi-Yau cone: The number of
internal points is the number of 4-cycles, and the number of 2-cycles
can be derived from the requirement that 
\#(0-cycles) + \#(2-cycles) + \#(4-cycles) = 2(Area); this
is the simply the Euler characteristic of the Calabi-Yau.
This number is also the same as the number of gauge groups in
the dual gauge theory. The Euler characteristic is
also the number of gauge groups in the dual quiver theory, as one
can check for these examples\footnote{This
rule was discovered empirically with B. Kol; the brane dimer picture
\cite{Hanany:2005ve,dimers} provides a proof.}.
Since each space here is connected, the number of 0-cycles is always 1.
It is then straightforward to figure out that the number of 2-cycles for
$\IF_0, dP_1$, and $dP_2$ is two, two, and three, respectively. This
corresponds exactly with out geometric intuition: Since 
$\IF_0 = \IP^1 \times \IP^1$, we expect two independent 2-cycles on $\IF_0$, and since
$dP_n$ is just $\IP^2$ blown up at $n$ points, we expect $n+1$ 2-cycles for
$dP_n$. 

\begin{figure}[ht]
  \epsfxsize = 10cm
  \centerline{\epsfbox{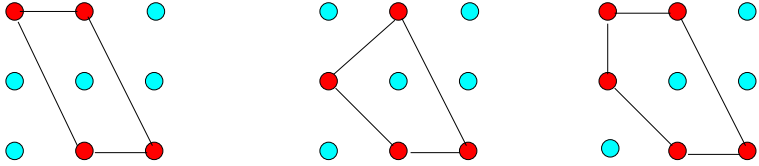}}
  \caption{Alternate toric data for $\IF_0$ (left) $dP_1$ (center), and $dP_2$ (right).}
  \label{toric2}
\end{figure}

Other (and for our purposes later on, more convenient) toric presentations of these
spaces are also possible. For future reference, we include these alternate presentations
in \fref{toric2}. One can easily check that these toric diagrams yield
the same number of 2- and 4-cycles as the ones in \fref{toric1}. The relation
between the two toric presentations is simply that the points along the 
diagonal in Figure \ref{toric1} have been brought to lie along a vertical line
in \fref{toric2}; the two toric diagrams correspond to different projections of 
the full three-dimensional toric diagram for the Calabi-Yau cones. The presentations
of \fref{toric1} can be mapped to those of \fref{toric2} by an affine transformation,
whose form we do not record here. 

Note that it is possible to see Higgsing from the toric diagram. Since
twice the area of the toric diagram (i.e. the number of triangles in 
a triangulation)
is the number of gauge groups in 
the dual superconformal theory, Higgsing corresponds to removing
an external point from the toric diagram. 
All the processes we consider here will never decrease the number of triangles
by more than one, meaning that we only consider Higgsings that
break $SU(N)^m \rightarrow SU(N)^{m-1}$ (cases for which this
is not true are more mysterious and not well understood). 
In the example above, it's easy to see how the toric diagram for
$dP_2$ can be altered to give the toric diagrams for $dP_1$ and $\IF_0$:
To get $dP_1$, delete the top left point from the $dP_2$ diagram in \fref{toric2};
to get $\IF_0$, delete the other point on the left side of the toric diagram. 
Note that the external lines in the toric diagrams correspond to places
where the $T^3$ fiber of the toric geometry has a degenerate cycle.
Removing an external line is thus the same as blowing down a 2-cycle;
this matches our intuition for how to get from $dP_2$ to $dP_1$ or $\IF_0$. 
It is also possible to see this process from the corresponding (p,q)-brane web,
which is straightforward to read off the toric diagram. 
We postpone the discussion of (p,q) webs until Section 4.

\begin{figure}[ht]
  \epsfxsize = 15cm
  \centerline{\epsfbox{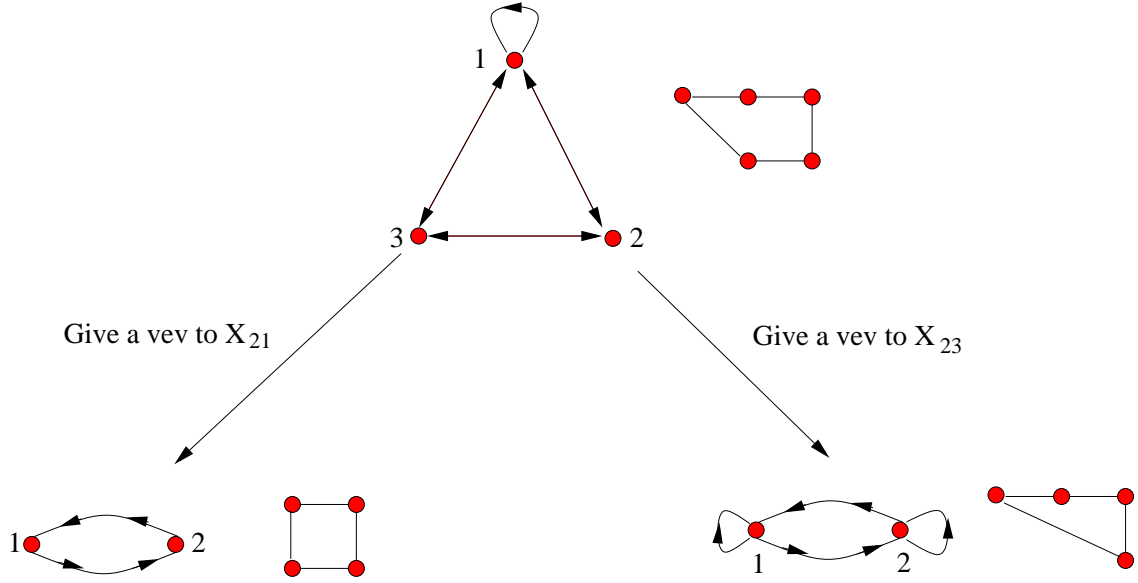}}
  \caption{The SPP quiver (top) can be Higgsed to the conifold (bottom left) or 
the $\IC^3/\IZ_2$ theory (bottom right).}
  \label{spp}
\end{figure}

As a final warm-up example, we recall the theory that blows down to the
gauge theories dual to the conifold and $S^5/\IZ_2$.
This theory is known as the Suspended Pinch Point (SPP), and was first described in 
\cite{Morrison:1998cs}. This theory can be Higgsed to the simplest $Y^{p,q}$ spaces, 
$Y^{1,0}$ and $Y^{1,1}$. $Y^{1,0}$ is simply the conifold theory, and
$Y^{1,1}$ is the ${\cal N}=2$ theory dual to $\IC^3/\IZ_2$. This Higgsing is
illustrated in \fref{spp}. 

The superpotential for the SPP is given by
\bea
W_{SPP}=X_{12}X_{23}X_{32}X_{21}-X_{23}X_{31}X_{13}X_{32}+X_{13}X_{31}X_{11}-X_{12}X_{21}X_{11}.
\label{wspp}
\eea
It is easy to see that upon setting $\langle X_{23} \rangle \neq 0$, the superpotential
becomes purely cubic. This is exactly as expected for the ${\cal N}=2$ theory, since
it is an orbifold of $\IC^3$. Giving $X_{21}$ a vev results in a mass term
for the fields $X_{12}$ and $X_{11}$, which must then be integrated out. Doing so
reproduces the superpotential for the conifold, given by two quartic terms. 

%=====================================================
\section{New quiver theories: $X^{p,q}$}
%=====================================================

In this section we give a procedure for constructing quivers which blow down
to $Y^{p,q}$ and $Y^{p,q-1}$.
The general idea is to start with a known quiver gauge theory, say $Y^{p,q}$, 
blow up its corresponding toric diagram and then to find the effect on the quiver. 
This procedure was done for several examples in \cite{Feng:2002fv}. In 
many situations there is a unique way to perform such a blow up; this gives the 
resulting quiver gauge theory. It turns out that in the case we study here 
we have such a situation in which a blow up gives a unique answer and therefore 
allows the construction of a new infinite family of quiver gauge theories.
We denote the quivers we construct in this paper by $X^{p,q}$; in this language, 
the $dP_2$ quiver would be called $X^{2,1}$, and the SPP
would be $X^{1,1}$. We will see that the procedure for constructing
$dP_2$ generalizes nicely to general $p$ and $q$. The blow-up we construct is
the unique possiblity for blowing up the specified node, but of course one could
always choose to blow up a different node. In a later section, we will show
that blowing up a different node simply results in a theory which is Seiberg dual to the one
presented here.

Before we continue on to the $X^{p,q}$ spaces, however, we briefly review the $Y^{p,q}$ quivers.
This will be a quick discussion; for more details the reader is referred to 
\cite{BFHMS}.

%=====================================================
\subsection{Review of the $Y^{p,q}$ quivers}
%=====================================================

\begin{figure}[ht]
  \epsfxsize = 10cm
  \centerline{\epsfbox{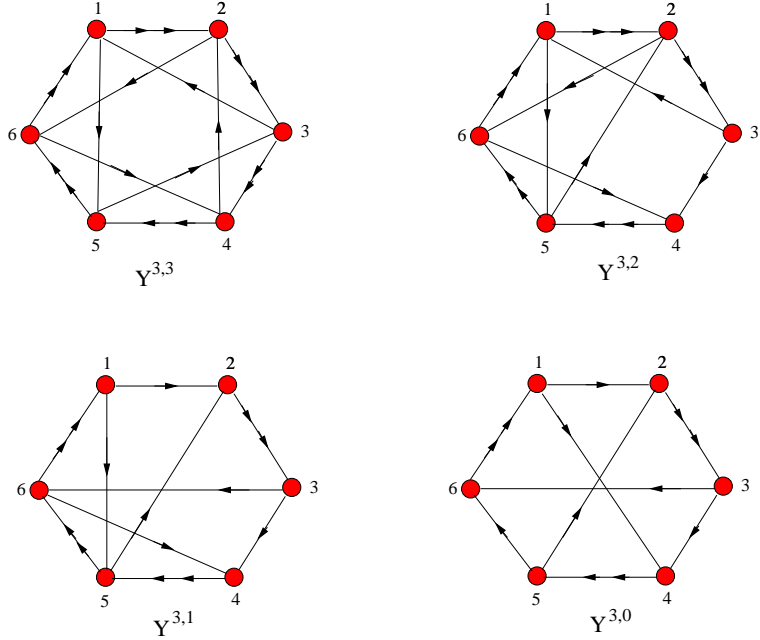}}
  \caption{Quivers for the $Y^{3,q}$ theories.}
  \label{y3q}
\end{figure}

The $Y^{p,q}$ theories were constructed in \cite{BFHMS}, following
the discovery of the dual geometries and
their toric descriptions in \cite{Gauntlett:2004yd,Gauntlett:2004zh,Martelli:2004wu}.  
It was shown there that they can be 
obtained as modifications of $Y^{p,p}$, which happens to be the theory living on a stack of D3 branes 
placed on the singular point of the orbifold  $\IC^3/\IZ_2 \times \IZ_p$.  
The orbifold theory can be constructed by standard methods as a projection of $\IN =4$ SYM.  The 
gauge group
is $SU(N)^{2p}$ where $N$  is the  number of D3 branes in the stack.  The $Y^{p,p}$ theory has 
$6p$ bifundamental
matter fields transforming in $2p$ doublets $U^{\alpha} , \alpha=1,2$, and $2p$ singlets 
$Y$ of the $SU(2)$ nonabelian part of the global symmetry group.
The superpotential for this theory descends from that of $\IN =4$ and  
consists of $4p$ cubic terms. The quiver
for $Y^{3,3}$  can be seen in the top left corner of \fref{y3q}. The superpotential for this theory is

\bea
 W_{Y^{3,3}} & = & \epsilon^{\alpha \beta} U_{12}^\alpha U_{23}^\beta Y_{31}
 +\epsilon^{\alpha \beta} U_{23}^\alpha U_{34}^\beta Y_{42} + \epsilon^{\alpha \beta} 
U_{34}^\alpha U_{45}^\beta Y_{53}\\ \nn
& + & \epsilon^{\alpha \beta} U_{45}^\alpha U_{56}^\beta Y_{64}
 +\epsilon^{\alpha \beta} U_{56}^\alpha U_{61}^\beta Y_{15} + \epsilon^{\alpha \beta} 
U_{61}^\alpha U_{12}^\beta Y_{26}. 
 \label{wy33}
 \eea
The doublets are contracted  in a way that respects the $SU(2)$ global symmetry. 

To construct the $Y^{p,p-1}$ theory,
one picks a doublet in $Y^{p,p}$, say the one between nodes $i$ and $i+1$, 
and removes one of the two bifundamentals. 
Then one removes the singlets $Y_{i+2,i}$, $Y_{i+1,i-1}$ and adds a new singlet $Y_{i+2,i-1}$.  
Four of the cubic 
terms in the superpotential are eliminated by these removals.  Finally, one adds two 
quartic terms to the superpotential,
involving the remaining of the two $U$ fields (now called $Z$), the new singlet and two $U$ doublets.
As an example, the quiver for $Y^{3,2}$ is shown in the top right corner of \fref{y3q}.  
The superpotential for this theory
reads:

 \bea
 W_{Y^{3,2}}& = & \epsilon^{\alpha \beta} U_{12}^\alpha U_{23}^\beta Y_{31}
 +\epsilon^{\alpha \beta}U_{23}^\alpha Z_{34}U_{45}^\beta Y_{52}\\ \nn
& + & \epsilon^{\alpha \beta} U_{45}^\alpha U_{56}^\beta Y_{64}
 +\epsilon^{\alpha \beta} U_{56}^\alpha U_{61}^\beta Y_{15} + \epsilon^{\alpha \beta} 
U_{61}^\alpha U_{12}^\beta Y_{26}. 
 \label{wy33}
 \eea 
 
This procedure repeated $p-q$ times for non-consecutive $U$ doublets yields $Y^{p,q}$. 
 In the lower 
half of \fref{y3q} we show the quivers for $Y^{3,1}$ and $Y^{3,0}$.  Each time $q$ 
descreases by one, four cubic terms 
are eliminated and two quartic terms appear in the superpotential. The superpotential of $Y^{p,q}$
therefore
has $4q$ cubics and $2(p-q)$ quartics. The modifications to the $Y^{p,p}$ quiver are called \emph{single 
impurities}; we say that there is a single impurity between any two nodes where there is a  
bifundamental $Z$.
The specific choice of sites on the quiver where single impurities are placed is not important, since the 
different theories 
obtained this way are related by Seiberg duality \cite{Benvenuti:2004wx}
and have the same IR dynamics. In fact, single impurities 
can 
be brought on top of each other and combine into \emph{double impurities}, which contribute cubic terms to 
the 
superpotential. We shall say more about these in a later section.  

%=====================================================
\subsection{$X^{p,q}$ for $1 \leq q \leq p-1$}
%=====================================================

\begin{figure}[ht]
  \epsfxsize = 15cm
  \centerline{\epsfbox{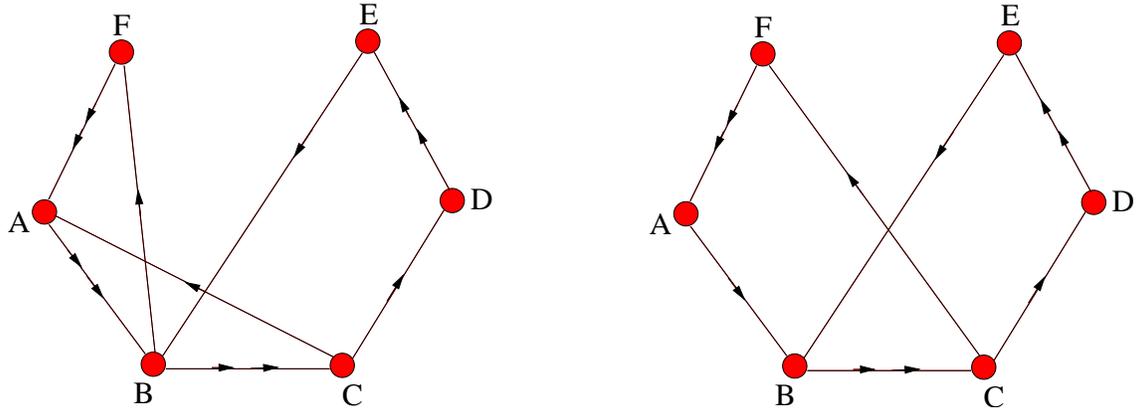}}
  \caption{The relevant portions of the quivers for $Y^{p,q}$ (left) and $Y^{p,q-1}$ (right).
 The rest of the quiver is assumed to be in between nodes E and F and is not drawn.}
  \label{quiver_ypqypq-1}
\end{figure}

Now that we have reviewed
the $Y^{p,q}$ theories, constructing the $X^{p,q}$ quivers is straightfoward. 
We first
consider the case $0 \leq q \leq p-1$, since the
procedure we give here must be altered slightly when $q=p$; this latter case will
be described subsequently.
Consider the quiver for $Y^{p,q}$. One toric phase of this theory will have
$p-q$ single impurities, where double arrows on the outside of the $Y^{p,p}$ 
quiver have been replaced
with single arrows.
 Since we only consider cases where $q \leq p-1$, there will always 
be at least one single arrow on the outside of the $Y^{p,q}$ quiver. Without loss
of generality, we can choose this arrow to be as close as possible to the 
 leg which is impurified when constructing the $Y^{p,q-1}$ quiver; if the single arrow is
 farther away, it is always possible to perform a sequence of Seiberg dualities to 
 bring it to the desired position. See \fref{quiver_ypqypq-1}.
 
 Let's call the node we blow up node $B$, which we blow up into the two nodes $B_1$ and $B_2$.
 Denote the node before $B$ by $A$, and the node after $B$ by $C$. 
 The new $X^{p,q}$ quiver is constructed as follows:
 \begin{itemize}
 \item Draw bifundamentals $X_{AB_1},X_{B_1B_2},X_{B_2C},X_{AB_2},X_{B_1C}$.
 \item For all single bifundamentals in the $Y^{p,q}$ quiver of the form $X_{nB}$ (i.e. entering $B$), draw a bifundamental $X_{nB_1}$.
 \item For all single bifundamentals in the $Y^{p,q}$ quiver of the form $X_{Bn}$ (i.e. exiting $B$), draw a 
 bifundamental $X_{B_2n}$.
 \item All other bifundamentals should be left as they are in the $Y^{p,q}$ quiver.
 \end{itemize}
 
\begin{figure}[ht]
  \epsfxsize = 6cm
  \centerline{\epsfbox{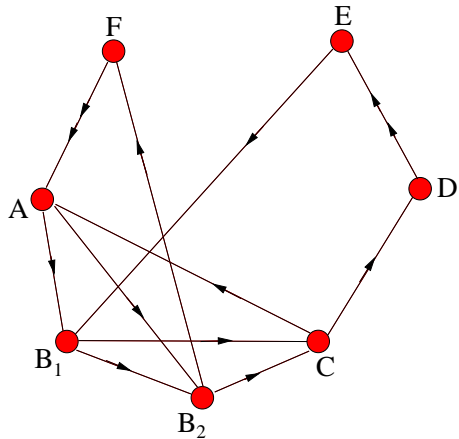}}
  \caption{The relevant portions of the quivers for $Y^{p,q}$ (left) and $Y^{p,q-1}$ (right).}
  \label{quiver_xpq}
\end{figure}

 For a graphical depiction of this process, see \fref{quiver_xpq}.

 Notice that there are $4p+2q$ bifundamentals in the $Y^{p,q}$ quiver. Four of them are the
 double arrows entering and exiting node $B$, which get replaced with five single arrows in
 the $X^{p,q}$ quiver. Thus there is a net increase of one in the number of arrows, meaning
 that this phase of our $X^{p,q}$ theories will always have $4p+2q+1$ bifundamentals. 
 As we saw in the previous section, this is exactly the case with $dP_2$,
 which has 11 fields, compared to the 10 fields of $dP_1$ and the 8 fields of $\IF_0$.
 
 Obtaining the superpotential is straightforward. We know that in the superpotential
 for a $Y^{p,q}$ quiver theory, there are $4q$ cubic terms and $2(p-q)$ quartic terms. To
 reproduce this upon Higgsing, we must have the superpotential
 \bea
 W_{X^{p,q}} & = & U^{(2)}_{DE}Y_{EB_1}Z_{B_1 B_2}Z_{B_2C}Z_{CD} +
 Z_{AB_1}Y_{B_1C}Y_{CA}+U^{(2)}_{FA}Y_{AB_2}Y_{B_2F} \\
 & - & Y_{EB_1}Y_{B_1C}Z_{CD}U^{(1)}_{DE} - Y_{AB_2}Z_{B_2C}Y_{CA} 
 -U^{(1)}_{FA}Z_{AB_1}Z_{B_1B_2}Y_{B_2F} \\
 & + & {\rm unchanged},
 \label{wxpq}
 \eea
 where ``unchanged" simply denotes all the other terms in the 
 original $Y^{p,q}$ superpotential, which are unaffected by the blowup. In this sense,
 blowing up a node is a ``local" procedure -- it only affects the fields within 3 nodes of the
 blown up node.
 
 That the $X^{p,q}$ quivers Higgs to $Y^{p,q}$ and $Y^{p,q-1}$ is now
easy to see. Setting  $\langle Z_{B_1B_2} \rangle \neq 0$ collapses
the nodes $B_1$ and $B_2$ back into node $B$. We lose the
field $Z_{B_1B_2}$ and the other fields remain, although
we should rewrite any $B_1$ and $B_2$ indices as $B$. This gives
(as it should, by our construction), the $Y^{p,q}$ quiver.
The superpotential (\ref{wxpq}) also does exactly what it must. Setting
 $\langle Z_{B_1B_2} \rangle \neq 0$ changes the quintic term into a quartic, and one
 of the quartics into a cubic. This gives the superpotential for $Y^{p,q}$, where it 
 is obvious that the global $SU(2)$ symmetry has been restored: The doublets
 are $(Z_{AB_1},Y_{AB_2}), (Y_{B_1C},Z_{B_2C}), (U^{(1)}_{FA},U^{(2)}_{FA}),$
 and $(U^{(1)}_{DE},U^{(2)}_{DE})$.
 
 Giving a vev to $Z_{B_2C}$ yields the quiver for $Y^{p,q-1}$. This affects the 
 superpotential in a mildly more nontrivial way than the previous case, since now
 the fields $Y_{AB_2}$ and $Y_{CA}$ get a mass and should be integrated out. Since
 these two fields appear in only cubic terms, the net effect of integrating them out
 is to replace the three cubic terms with one quartic. This quartic
 is exactly what we'd expect; it is paired with $U^{(1)}_{FA}Z_{AB_1}Z_{B_1B_2}Y_{B_2F}$
 under the newly restored $SU(2)$ symmetry. For completeness, we note that
 the new $SU(2)$ doublets are given by $(U^{(1)}_{FA},U^{(2)}_{FA}), (U^{(1)}_{DE},U^{(2)}_{DE}),$
 and $(Z_{B_1B_2},Y_{B_1C})$.
 
 We note here that the total number of terms of a given degree in the 
 superpotential for $X^{p,q}$ is easy to figure out. There is always one quintic, 
 $2(p-q)$ quartics, and $4q-1$ cubic terms. The reason is clear: In the $Y^{p,q}$ quiver,
 there are $2(p-q)$ quartics and $4q$ cubics. Since blowing down
 the $X^{p,q}$ quiver to $Y^{p,q}$ involves shifting a quintic to a quartic and a quartic to a 
 cubic, we see that the number of quintic terms
in the $X^{p,q}$ superpotential is one,  whereas the net number of cubic terms decreases by 
one and the net number of quartics remains the same. 

%=====================================================
\subsection{$X^{p,q}$ for $q=p$}
\begin{figure}[ht]
  \epsfxsize = 6cm
  \centerline{\epsfbox{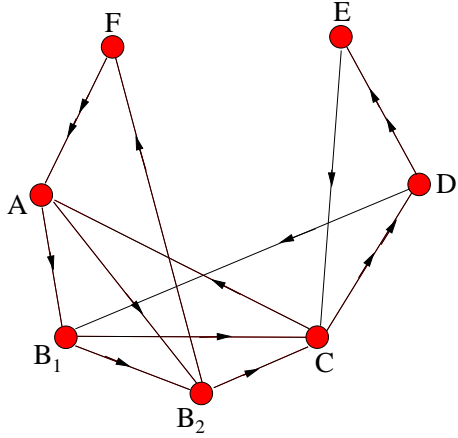}}
  \caption{The relevant portions of the quivers for $Y^{p,q}$ (left) and $Y^{p,q-1}$ (right).}
  \label{quiver_xpp}
\end{figure}

%=====================================================
Now, let's consider the case $q=p$. The above procedure clearly must be modified, 
since a quintic term in the superpotential may no longer exist since there is nothing
for it to descend to in the $Y^{p,p}$ theory. Nevertheless, the procedure is more or less the 
same as above, the only difference being that instead of drawing a bifundamental
between node $E$ and node $B_1$, we draw one between $D$ and $B_1$. This is 
shown in \fref{quiver_xpp}; as before, the parts of the quiver that
do not change are not shown. 

The superpotential is now given by

 \bea
 W_{X^{p,q}} & = & Z_{B_2 C}Y_{CA}Y_{AB_2}+U_{CD}^{(2)}Y_{DB_1}Y_{B_1C}
 -Y_{B_1C}Y_{CA}Z_{AB_1}-Y_{AB_2}Y_{B_2F}U^{(2)}_{FA} \\
 & - & Z_{B_2C}U^{(1)}_{CD}Y_{DB_1}Z_{B_1B_2}+Z_{B_1B_2}Y_{B_2F}U^{(1)}_{FA}Z_{AB_1} \\
 & + & {\rm unchanged},
 \label{wxpP}
 \eea
 
 Notice that, as before, when we set $\langle Z_{B_1B_2} \rangle \neq 0$, the theory
 flows to $Y^{p,p}$, and when we set $\langle Z_{B_1C} \rangle \neq 0$, the theory
 flows to $Y^{p,p-1}$. In the latter case, the fields $Y_{AB_2}$ and $Y_{AC}$ acquire
 a mass and should be integrated out; this yields the correct superpotential
 for $Y^{p,p-1}$. 
 
 For general $p$, then, we see that this $X^{p,p}$ theory has $6p+1$ fields. The superpotential
 has 2 quartic terms and $4p-2$ cubics. Going to the $Y^{p,p}$ theory simply
 changes both quartics into cubics, which recovers the $4p$ cubic terms
 required for this theory. Flowing to the $Y^{p,p-1}$ theory involves
 shifting one quartic to a cubic, and taking three cubics into one quartic. Thus, the
 resulting theory has two quartic terms and $4(p-1)$ cubics, which is correct
 for $Y^{p,p-1}$.
 
 It is worth pointing out again that the above prescription gives merely one
 way of constructing the $X^{p,q}$ theories, and there are many 
 different possible toric phases of these theories. We will explore these 
 dualities in a later section.

%===========================================================
\subsection{An example: $X^{3,1}$, $X^{3,2}$, and $X^{3,3}$}
%===========================================================

\begin{figure}[ht]
  \epsfxsize = 10cm
  \centerline{\epsfbox{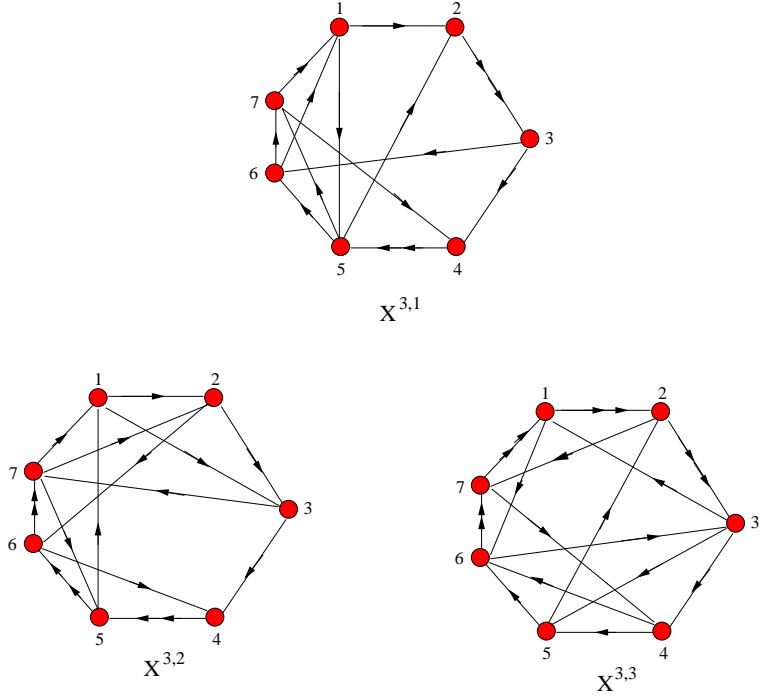}}
  \caption{Quivers for a particular phase of $X^{3,1}$, $X^{3,2}$, and $X^{3,3}$.}
  \label{x3q}
\end{figure}

As an illustrative example, we now present quivers for $X^{3,1}$, $X^{3,2}$, and 
$X^{3,3}$. These theories will Higgs to the quivers for 
$Y^{3,0}, Y^{3,1},Y^{3,2},$ and $Y^{3,3}$ in \fref{y3q}.
Apply the procedure outlined in the previous sections is straightforward, and
yields the quivers in \fref{x3q}.

We can easily write down the superpotentials
for these theories. They are

 \bea
 W_{X^{3,1}} & = &
 U_{23}^{(2)}Y_{36}Z_{67}Z_{71}Z_{12} + Z_{56}Y_{61}Y_{15} + U_{45}^{(1)}Y_{57}Y_{74} \\
\nn  & - & Y_{36}Y_{61}Z_{12}U_{23}^{(1)} - Y_{57}Z_{71}Y_{15} - U_{45}^{(2)}Z_{56}Z_{67}Y_{74}
 \\
\nn  & + &  \epsilon^{\alpha \beta} U_{23}^\alpha Z_{34}U_{45}^\beta Y_{52}, 
 \label{wx31}
 \eea
 
  \bea
 W_{X^{3,2}} & = &  U_{45}^{(2)}Y_{51}Z_{12}Z_{23}Z_{34} + Z_{71}Y_{13}Y_{37} + 
U_{67}^{(1)}Y_{72}Y_{26} \\
\nn  & - & Y_{51}Y_{13}Z_{34}U_{45}^{(1)} - Y_{72}Z_{23}Y_{37} - 
U_{67}^{(2)}Z_{71}Z_{12}Y_{27} \\
\nn & + &  \epsilon^{\alpha \beta} U_{45}^\alpha U_{56}^\beta Y_{64}
 +\epsilon^{\alpha \beta} U_{56}^\alpha U_{67}^\beta Y_{71}, 
 \label{wx32}
 \eea
 
and

  \bea
 W_{X^{3,3}} & = &
 Z_{56}Y_{63}Y_{35}+U_{67}^{(2)}Y_{74}Y_{46}
 -Y_{46}Y_{63}Z_{34}-Y_{35}Y_{52}U^{(2)}_{23} \\
\nn & - & Z_{56}U^{(1)}_{67}Y_{74}Z_{45}+Z_{45}Y_{52}U^{(1)}_{23}Z_{34} \\
 \nn & + & \epsilon^{\alpha \beta} U_{67}^\alpha U_{71}^\beta Y_{16}
 +\epsilon^{\alpha \beta} U_{71}^\alpha U_{12}^\beta Y_{27} + \epsilon^{\alpha \beta} 
U_{12}^\alpha U_{23}^\beta Y_{31}. 
 \label{wx33}
 \eea

As they must, these become the $Y^{3,q}$ superpotentials
upon giving vevs to the appropriate fields and integrating out where
necessary.

%===========================================================
\subsection{Toric Diagrams for $X^{p,q}$}
%===========================================================

\begin{figure}[ht]
  \epsfxsize = 10cm
  \centerline{\epsfbox{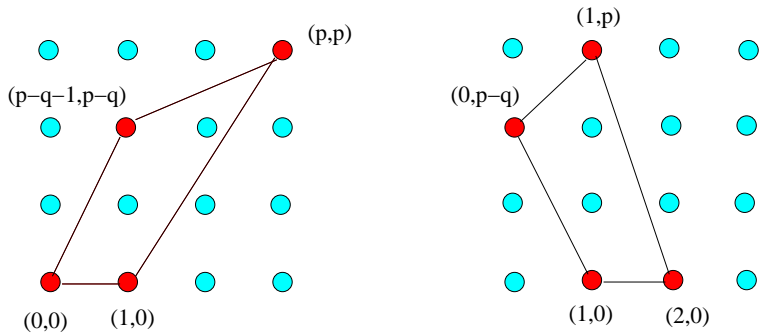}}
  \caption{Two different presentations of the toric data for $Y^{p,q}$. The
  figures here are $Y^{3,1}$. }
  \label{toricypq}
\end{figure}

We now describe the toric presentation of the $X^{p,q}$ spaces. As discussed in 
the introduction, it is a rare occurrence to know the toric diagram
corresponding to a given quiver. The Forward Algorithm \cite{Feng:2000mi} can be used to 
extract the toric data for simple quivers, but it is computationally
prohibitive for quivers with many nodes. So knowing the toric
data dual to an infinite number of quivers is highly nontrivial. 
Finding the toric data for
the $X^{p,q}$ theories is straightforward, 
since we can use our intuition from $dP_2$ to simply
write down the answer and then check that it is correct. First, we note
two different toric presentations of the space $Y^{p,q}$; these are given
in \fref{toricypq}. The toric data on the left is the presentation used in
\cite{Martelli:2004wu,BFHMS}; the toric data on the right is simply an alternate projection which 
is particularly useful for our purposes\footnote{This
representation was also used in \cite{Franco:2005fd}.}. Notice that for a given $p$, the only point that
moves is the point along the left edge of the lattice. As $q$ decreases, this point
moves {\bf up}; at $q=0$ we recover the expected parallelogram for $Y^{p,0}$. 

\begin{figure}[ht]
  \epsfxsize = 5cm
  \centerline{\epsfbox{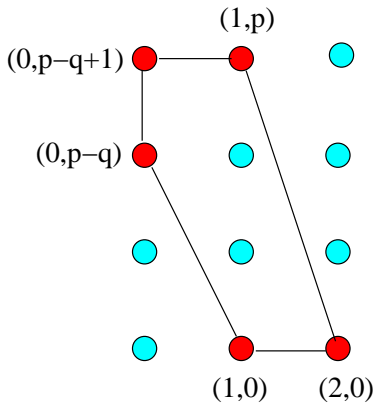}}
  \caption{The toric data for $X^{p,q}$. This diagram is for $X^{3,1}$,
  which blows down to $Y^{3,0}$ and $Y^{3,1}$.}
  \label{toricxpq}
\end{figure}

The toric data for $X^{p,q}$ is now easy to intuit. Since we need a space which 
blows down to both $Y^{p,q}$ and $Y^{p,q+1}$, we simply include both points on 
the left edge of the lattice, as in \fref{toricxpq}. Removing the top left point
leaves the toric diagram for $Y^{p,q}$, and removing the one below it
yields the toric diagram for $Y^{p,q-1}$. 
Note that, as for 
$Y^{p,q}$, these two points move ${\bf up}$ as $q$ decreases.

There are many checks that these are the correct toric diagrams dual to the
$X^{p,q}$ quivers. First, one may use the Forward Algorithm of \cite{Feng:2000mi} to
extract the toric data. These have been checked for small $p$, and yield the
expected results \footnote{We thank Alan Dunn for this calculation.}. We may also
check the number of gauge groups, which is equal to the number of triangles in a triangluation
of the toric diagram. This should be equal to $2p+1$ in general, and it is straightforward
to see that this is correct: The $Y^{p,q}$ theories have $2p$ triangles, and the
effect of adding the extra node is to add one more. Additionally, we can
read off the even Betti numbers of the Calabi-Yau cone. There are $p-1$ internal
points, corresponding to $p-1$ 4-cycles, and since the space
is connected there is only one 0-cycle. Therefore, there
must be $p+1$ 2-cycles. Thus, the number of 2-cycles for the $X^{p,q}$ theories
is one greater than it is in the lower $Y^{p,q}$ theories,
as we expect by analogy to $dP_2$.

Notice that we can rephrase the above as follows: If $I$ is the number of internal points
in the toric diagram and $E$ is the number of external points, then
the number of 4-cycles is $I$, and the number of 2-cycles is $I + E -3$. This has an
interpretation in the related five dimensional gauge theory, as we will discuss in the next section.

We also can find some properties of the corresponding Sasaki-Einstein manifold
at the base of the Calabi-Yau cone\footnote{
We thank James Sparks for discussions on this.}. For a Sasaki-Einstein space
whose toric diagram has
$d$ external lines, $H_3= \IZ^{d-3}$, so here we find $H_3(X^{p,q}) = \IZ^2$.
Thus there are always two 3-cycles available for a D3-brane to wrap; this will be
discussed further in Section 6. The topological possibilities for
five-dimensional spaces are known, thanks to Smale's Theorem. Here,
we can say that the $X^{p,q}$ Sasaki-Einstein manifolds for $p \neq q$ (the
$p=q$ case is singular) are a
connected sum $(S^2 \times S^3)\#(S^2 \times S^3)$.

Knowing that the toric diagram in \fref{toricxpq}
gives a Calabi-Yau dual to the $X^{p,q}$ theories is highly nontrivial. Although
our construction of the gauge theories is done without ever considering
the dual geometry, it is important to point out that we know information about
{\bf both} sides of the AdS/CFT duality. The metrics on the $X^{p,q}$ geometries
are not known, and appear to be quite difficult to calculate. The
$Y^{p,q}$ theories had a global $SU(2)$ symmetry;
the existence of this non-Abelian symmetry was crucial to figuring out
the metrics on the Sasaki-Einstein manifolds \cite{Gauntlett:2004yd}. The $X^{p,q}$ gauge theories
have only $U(1)$ global symmetries, so we lose the power of the non-Abelian isometry 
when trying to find the dual metrics. Thus, the Sasaki-Einstein metrics on these spaces
are unknown, and probably rather difficult to derive.

\section{(p,q)-brane webs and 5d gauge theories}

It is known that one can get a five-dimensional theory associated to the
theory living on the D3-branes at the tip of the Calabi-Yau cone by writing
down a web of (p,q)\footnote{Here we run into the
problem of using the grouping (p,q) in two different contexts. (p,q) for 5-branes
will always be in Roman, and $(p,q)$ for $Y^{p,q}$ will be in
math (italic).} 5-branes \cite{Aharony:1997ju}; 
the procedure for deriving the brane web from the corresponding toric diagram 
is well-known \cite{Leung:1997tw,Aharony:1997bh}. More mysterious, however, is
what one can say about the resulting 5-dimensional gauge theory living on the
4+1 dimensions common to all the 5-branes. Some things are known about the
correspondence between the five dimensional theory and the related
four dimensional quiver \cite{Franco:2002ae}, but much still remains unkown. Let us now
briefly review what is known about (p,q)-webs and the associated 5 dimensional
gauge theories.

Type IIB string theory has a vast armada of 5+1 dimensional branes, the (p,q) 5-branes.
(p,q) 5-branes are bound states of different numbers of D5-branes and NS5-branes;
we take the convention that a (1,0) brane is a D5-brane, and a (0,1) brane is an NS5-brane.
A (p,q) 5-brane is simply the magnetic dual of a (p,q) string, and may be thought
of as coming from an M5-brane wrapped on a (p,q)-cycle of a $T^2$.
The tension of an arbitrary (p,q) brane is then given by $T_{(p,q)} = |p + \tau q|T_{D5}$, 
where $\tau$ is the Type IIB axion-dilaton. These (p,q) 5-branes 
are useful tools for studying 5d gauge theories, via arranging the branes
in a network such that they share 4+1 dimensions. The remaining dimension can be taken
to lie in a plane, and the branes can be arranged in a network called a (p,q) web. 
Of course, placing branes at generic angles will break all supersymmetry. The requirement
that a web be stable and preserve supersymmetry can be summed up in the conditions
\bea
\frac{\Delta x}{\Delta y} = \frac{p}{q} \qquad {\rm and} \qquad \sum_i p_i = \sum_i q_i =0.
\label{stableweb}
\eea
In (\ref{stableweb}), the first condition states that the slope of a brane in the $(x,y)$ plane is
equal to the ratio of its (p,q) charges, and the second condition 
is simply conservation of p and q charge at each vertex (the sum is
over all branes ending at a given vertex). The slope condition ensures that one quarter of
the supersymmetry is preserved, giving the 8 supercharges for a five-dimensional
${\cal N} =1$ theory.  

\begin{figure}[ht]
  \epsfxsize = 12cm
  \centerline{\epsfbox{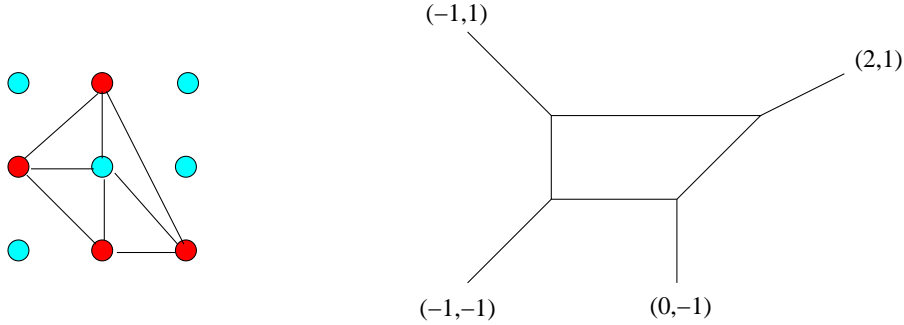}}
  \caption{A triangulation of the toric data for $dP_1$ and the corresponding brane web.}
  \label{dp1triang}
\end{figure}

It is now well-known that one can associate a toric diagram to a (p,q) web via
a straightforward procedure, which we
now review. First, one needs to pick a triangulation of the toric diagram;
we have done this for $dP_1$ in \fref{dp1triang}. The brane web is now essentially
just the dual of this diagram: To construct it, just draw the lines orthogonal to the lines in
the triangulated toric diagram. External lines in the toric diagram
correspond to semi-infinite branes, and internal lines correspond to branes with finite extent
in the $(x,y)$-plane. Notice that a consequence of this is that 
the number of internal points in the toric diagram 
corresponds to the number of closed polygons in the brane web.

\begin{figure}[ht]
  \epsfxsize = 12cm
  \centerline{\epsfbox{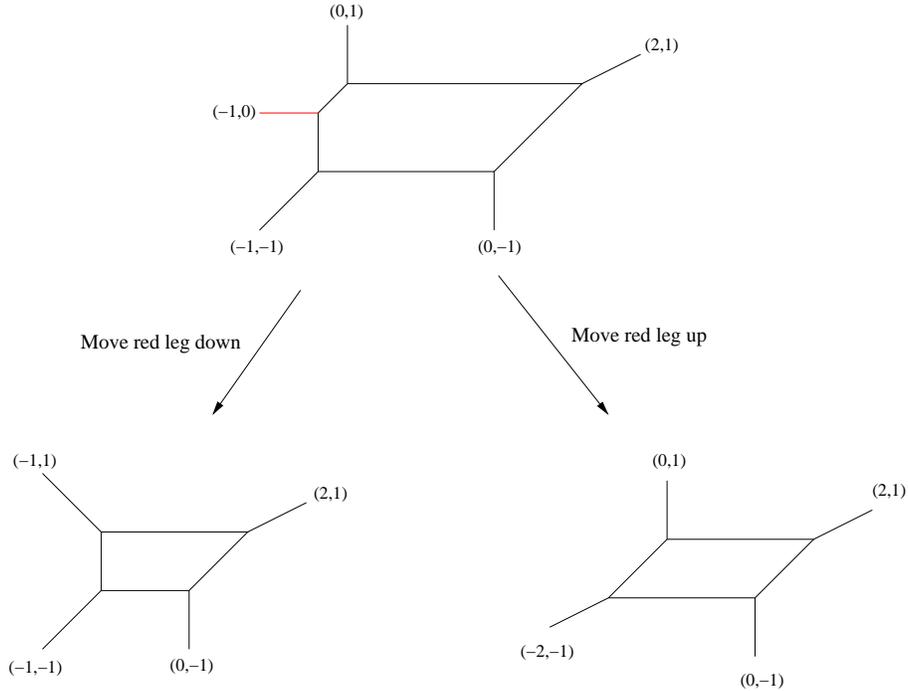}}
  \caption{By moving the horizontal (red) brane in the above $dP_2$ brane web
  up or down, one can get a brane web for $dP_1$ (left) or $\IF_0$ (right).}
  \label{dp2dp1f0}
\end{figure}

As a warm-up for finding the $X^{p,q}$ brane webs, let's see how one can use the
web for $dP_2$ to get the associated webs for $dP_1$ and $\IF_0$. Related discussion can be found near Figure 12 
of \cite{Aharony:1997ju}. The brane webs for $dP_2$, $dP_1$, and $\IF_0$ are
shown in \fref{dp2dp1f0}; we have used the toric data of \fref{toric2} to
construct them. Notice that the $dP_2$ web has a semi-infinite horizontal
brane, which we have drawn in red. This is a D5-brane, and shows up in the 5d
theory as a massive flavor. As we move this brane up or down, the mass of the flavor
changes and it may then be integrated out of the theory. Interestingly, the resulting theory
is different depending how one increases the mass: By moving the (-1,0) D5-brane up, it hits
the (0,1) brane and results in a (-1,1) brane, giving the web for $dP_1$. By moving the
D5-brane down, it hits the (-1,-1) brane at the bottom and results in a (-2,-1) brane, giving 
the $\IF_0$ brane web. We will see analogous behavior in the $X^{p,q}$ brane webs when
we generate the two possible descendant $Y^{p,q}$ webs. 

\begin{figure}[ht]
  \epsfxsize = 15cm
  \centerline{\epsfbox{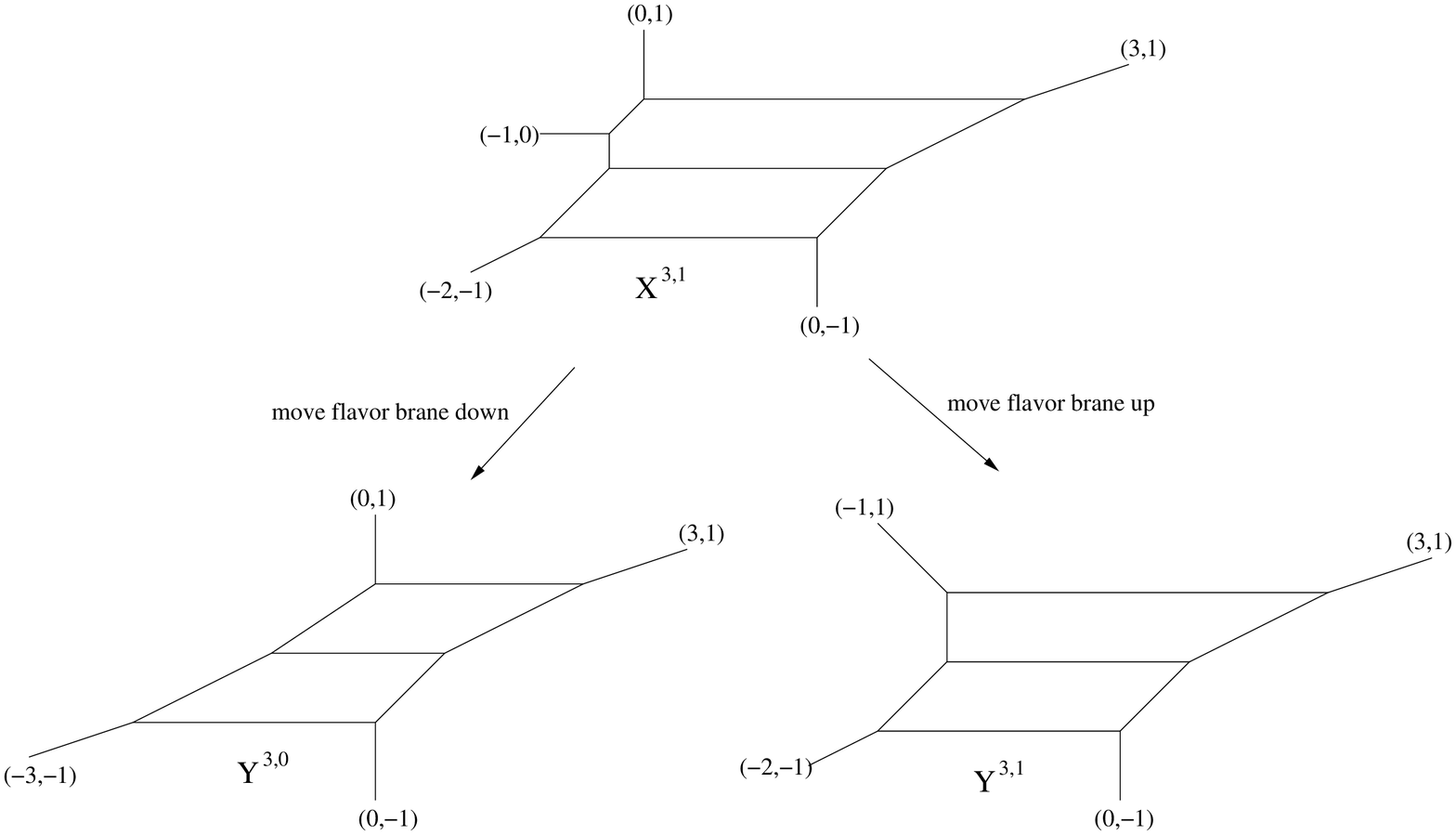}}
  \caption{The process of going from the $X^{3,1}$ web to that of $Y^{3,0}$ or $Y^{3,1}$.}
  \label{xpqweb}
\end{figure}

We now see that knowing the toric diagram is tantamount to knowing the brane web. One can
easily see in \fref{toricxpq} that we'll always get a flavor D5-brane which can
be moved up or down; this is just the external line dual to the one external vertical line
in the toric diagram. 
We do note that the situation is mildly more complicated for many internal points, since 
moving the flavor D5 past any brane junction means that one must change the (p,q) charges of
the branes at the junction. One example of this procedure is done, for $X^{3,1}$, in
\fref{xpqweb}.

We also note here that integrating out the massive flavor is equivalent, geometrically, to
blowing down a 2-cycle. This is especially simple to see in the $dP_2$ example, since
we know that $dP_2$ blows down to either $dP_1$ or $\IF_0$. This
is also easy to see from the toric diagram, since
external lines correspond to 2-cycles. For the
general
$X^{p,q}$ theoriesthis interpretation of
blowing down a 2-cycle remains true. For
an interesting discussion of the transitions between
some of the $X^{p,q}$ and $Y^{p,q}$ theories, see Figure 25 of \cite{Aharony:1997bh} 
and the surrounding discussion.

The $dP_2$ theory, or $X^{2,1}$ as we call it in this paper, 
corresponds to a five dimensional $SU(2)$ gauge theory with one flavor. 
Taking the mass of this flavor to positive infinity yields one of two 
different theories: If we define the sign of the mass to correspond to the 
positive $y$ direction in \fref{dp2dp1f0}, this theory is $\IF_0$.
Taking the mass to negative infinity gives the other theory, 
 $dP_1$. The distinction between these two theories is related to the value of the 
discrete $\theta$ angle, which follows from the fact that the
4th homotopy group of the gauge group is 
$\pi_4(SU(2))=\IZ_2$. This distinction is special to the case of $SU(2)$ as this 
is the only unitary group with a non-trivial 4th homotopy group. For higher values 
of the rank of the unitary group the five dimensional theory can admit a Chern-Simons 
term. This term is absent for the $SU(2)$ case since the completely symmetric
rank 3 invariant for $SU(2)$ vanishes, while for the other $SU(p)$ theories it is 
non-vanishing. As a result, for the 5d theories
living on the brane webs dual to the $Y^{p,q}$ toric diagrams with
$p>2$ we can introduce a Chern-Simons term which 
turns out to have a coefficent equal to $q$.
As discussed in detail in \cite{Intriligator:1997pq} 
a massive flavor contributes at the one loop level to the effective Chern-Simons coupling 
a value of $1/2$ and the sign of this contribution is proportional to the sign of the 
mass of this flavor. As a result when going from a large positive mass to a large negative 
mass and vice versa, the value of the Chern-Simons coefficient changes by 1. As we 
identify the Chern-Simons coupling with $q$, taking the mass from one sign to another 
precisely maps to the process of changing $q$ by 1. This is the process we have discussed
above in which we start with $X^{p,q}$, corresponding to a small mass in the 5d theory, and
Higgs to either $Y^{p,q}$ or $Y^{p,q-1}$ by giving a large positive or negative mass, respectively,
to the 5d flavor.

One more interesting point to note about the $Y^{p,q}$ theories is that from a 
five dimensional point of view, the allowed values for $p$ and $q$ which give interacting UV 
fixed points in five dimensions are $p>q\ge0$.
The case with $q=0$ is the simplest five dimensional SYM $SU(p)$ gauge theory 
where there is no Chern-Simons term. The case with $q=p$ does not lead to an 
interacting UV fixed point since the brane web involves parallel legs; see a 
discussion in \cite{Aharony:1997bh}. The conditions which were written in 
\cite{Intriligator:1997pq} for an interacting fixed point coincide nicely 
with the allowed range of the $Y^{p,q}$ theories as required by the geometry side to have smooth metrics.
These conditions can be extended to the case of more flavors. If we denote 
the number of flavors of the five dimensional gauge theory by $N_f$ then this 
number is related to the number of external nodes $E$ in the corresponding 
toric diagram by $E=N_f+4$.  The condition for an interacting UV fixed point is then
$N_f+2q<2p$. For $N_f=0$ this recovers the known limits on $Y^{p,q}$ discussed 
above while for $N_f=1$ we get a new bound which is consistent with the limits 
that we have found in this paper.
We further get a prediction for the allowed ranges of theories for higher values of $N_f$.

As discussed in \cite{Aharony:1997bh}, for a given five dimensional gauge theory 
with $SU(p)$ gauge group and for any Chern-Simons coefficient $q$ the number of 
parameters in the Lagrangian is the number of external legs $E$ in the
(p,q) web minus 3. As 
an example, for $N_f=0$ the number of parameters is 1; this parameter is simply
the gauge 
coupling of the five dimensional gauge theory. For $N_f=1$ there is an additional 
parameter given by the mass of the flavor, etc. This number $E-3$ actually counts 
the number of baryonic $U(1)$ global symmetries in the corresponding quiver gauge 
theory. Thus for the $Y^{p,q}$ theories we have one $U(1)_B$ and for $X^{p,q}$ 
we expect two baryonic $U(1)$ global symmetries.

There are additional matchings we can make between the five dimensional theory and 
the quiver gauge theory. The number of moduli for the five dimensional theory is 
equal to the rank of the gauge group, $p-1$. This number gives the number of distinct 
monopole solutions in the five dimensional gauge theory as well as the various vacuum 
expectation values
which can spontaneously break gauge invariance. For the geometry 
this number is the number of internal points in the toric diagram and is therefore the 
number of vanishing 4-cycles for the singular geometry. The number of 2-cycles in the 
singular geometry has yet another simple expression as $p-1+E-3$. This number also counts 
the number of different BPS charges $B$ that particles can carry in the five dimensional 
theory. As is well known the number of gauge groups for the quiver gauge theory is given 
by the total number of all even (0-,2-,and 4-) dimensional cycles. Therefore we get a relation 
which states that the number of different monopole solutions, denoted $M$, is given in 
terms of $p$ and the number of external lines as $M+B+1 = 2M+N_f+2 = 2M+E-2 = 2p+N_f$. 
For the cases $N_f=0$ and $N_f=1$ we recover the known cases of $Y^{p,q}$ and $X^{p,q}$, 
respectively. See related discussions in \cite{Aharony:1997bh}.
We summarize the above discussion in Table \ref{4d5d}.

\begin{table}[!h]
\begin{center}
$$\begin{array}{|c|c|}  \hline
\mathrm{5d} \,\, $SU(p)$ \,\,\mathrm{theory} & Y^{p,q}\,\, \mathrm{Toric}\,\,\mathrm{Geometry}
\\ \hline\hline
\mathrm{Number\,\, of\,\, Monopoles\,\,} $M$ & p-1  \\\hline
\mathrm{Number\,\, of\,\, BPS\,\,States\,\,} $B$ &  \mathrm{Number\,\,of\,\,2-cycles,\,\,}
p-1+E-3 \\\hline
\mathrm{5d\,\, Moduli} &  \mathrm{Number\,\,of\,\,4-cycles,\,\,} p-1 \\\hline
\mathrm{Chern-Simons\,\, Coefficient\,\,} & q   \\\hline
\mathrm{Number\,\, of\,\, Flavors\,\,} N_f & E-4  \\\hline

\end{array}$$
\caption{The relationship between different parameters in the 5d theories and $Y^{p,q}$ toric
geometries.}
\label{4d5d}
\end{center}
\end{table}
%===========================================================
\section{Toric Phases of the $X^{p,q}$ theories}
%===========================================================
The $X^{p,q}$ quiver gauge theories we present in this paper each flow to
a superconformal fixed point at the infrared.
At that point one can apply Seiberg duality \cite{Seiberg:1994pq} to any of the gauge
groups and get a new theory which has a different matter content
and superpotential but flows to the same IR fixed point as the original
theory.  By repeating this process it is possible to
obtain an infinite number of UV inequivalent theories that fall into the same
universality class.  In this section we will discuss a
particularly simple finite subclass of these theories, the {\it connected
toric phases}. We define the term {\it toric phase} to mean
that all the gauge groups have the same rank.  
The term ``connected'' refers to the theories generated
by starting from a toric phase, like the ones discussed above, and only
dualizing nodes with number of flavors ($N_f$) equal to twice the
number of colors ($N_c$). These are sometimes called \emph{self-dual}
nodes, because the
gauge group is unchanged by the dualization.  
The resulting gauge theory is again in a toric phase, with gauge
group
$SU(N)^{2p+1}$ and every field appearing in the superpotential exactly
twice.
For the sake of brevity we will omit the tedious details of the
dualizations.  Instead we shall outline the general structure of
these toric phases, using a few of the results of  \cite{Benvenuti:2004wx}
for the corresponding $Y^{p,q}$ quivers. 

\begin{figure}[ht]
 \epsfxsize = 8cm
 \centerline{\epsfbox{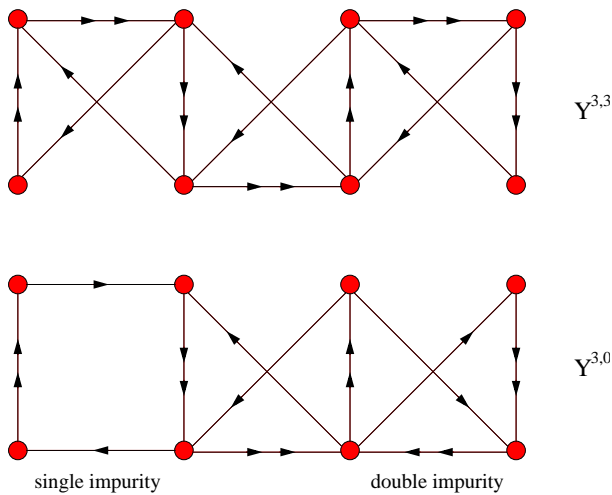}}
 \caption{The quivers for $Y^{33}$ and $Y^{30}$ with one single (left block) and one double impurity (right block).}
 \label{ypqexample}
\end{figure}

The toric phases of the $Y^{p,q}$ theories can be described as  modifications
of the quiver for $Y^{p,p}$.  These can be seen very clearly in the ``ladder''
depiction of the quiver, in which the nodes are placed in two
parallel rows  and numbered in a crenellating\footnote{{\it crenellating}, adj.:
having repeated square indentations like those in a battlement.}
 fashion. The quiver is then made up
of ``blocks,'' i.e. square sections between rungs
of the ladder.  All the blocks are identical in $Y^{p,p}$ and each one can be replaced
by a single or double impurity.  The numbers of single impurities ($n_1$) and  double
impurities ($n_2$) are restricted by the relation $n_1+2n_2=p-q$.  An example of this 
construction is shown in \fref{ypqexample}.

In these toric phases of the $Y^{p,q}$ theories, the only self-dual nodes
are the ones at the ends of a single or double impurity.
The blow-up procedure which produces the $X^{p,q}$ theories creates new
self-dual nodes. More precicely, node $B$ in $Y^{p,q}$
\fref{quiver_ypqypq-1} is
replaced by self-dual nodes $B_1$ and $B_2$ in $X^{p,q}$
\fref{quiver_xpq}. Dualizing gauge groups that do not share any
bifundamentals with these two nodes has
exactly the same effect as in the case of $Y^{p,q}$.  The impurities of
$Y^{p,q}$ can be moved around the quiver, fusing into double impurities
when
they collide.  As an example, we  show the result of dualizing node 6 of
$X^{3,1}$ below.  The two impurities fuse into a double impurity
exactly as they would in the absence of the blow up. 
Dualizing node 7 has an essentially identical effect.

\begin{figure}[ht]
 \epsfxsize = 8cm
 \centerline{\epsfbox{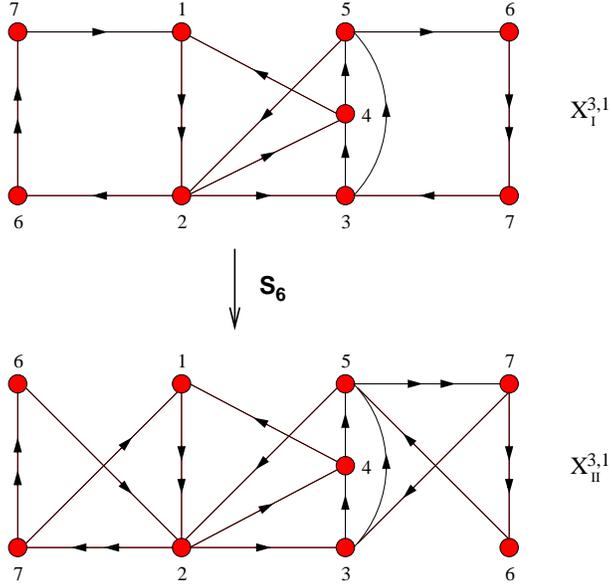}}
 \caption{The notation $S_6$ means Seiberg duality on node 6. The
resulting double impurity is located between nodes 1 and 6.}
 \label{doubleimpurity}
\end{figure}

\begin{figure}[ht]
 \epsfxsize = 6.8cm
 \centerline{\epsfbox{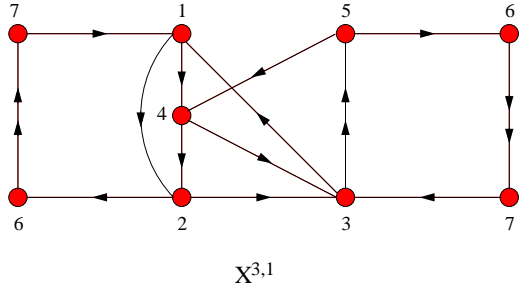}}
 \caption{Seiberg duality on node 4 corresponds to blowing up a different
node of $Y^{3,1}$.}
 \label{dualize4}
\end{figure}

On the other hand, dualizing nodes connected by an arrow to the new
self-dual nodes leads to new theories, not always accounted for by the
toric phase structure of $Y^{p,q}$.  Again we will use $X^{3,1}$ as a
showcase. Dualizing node 4 leads to the quiver shown in
\fref{dualize4}.  We have moved  node $4$ between nodes 1 and 2 to make
clear the result of this dualization:  The theory we
get is the same as the one we would get by blowing up node 2 instead of 3,
as mentioned in Section 3.  Higgsing $X_{14}$ in this quiver gives $Y^{3,1}$ with
two single impurities, while Higgsing $X_{42}$ gives $Y^{3,0}$ with three
single impurities. Dualizing 5 gives a theory completely
equivalent to this one.

\begin{figure}[ht]
 \epsfxsize = 14cm
 \centerline{\epsfbox{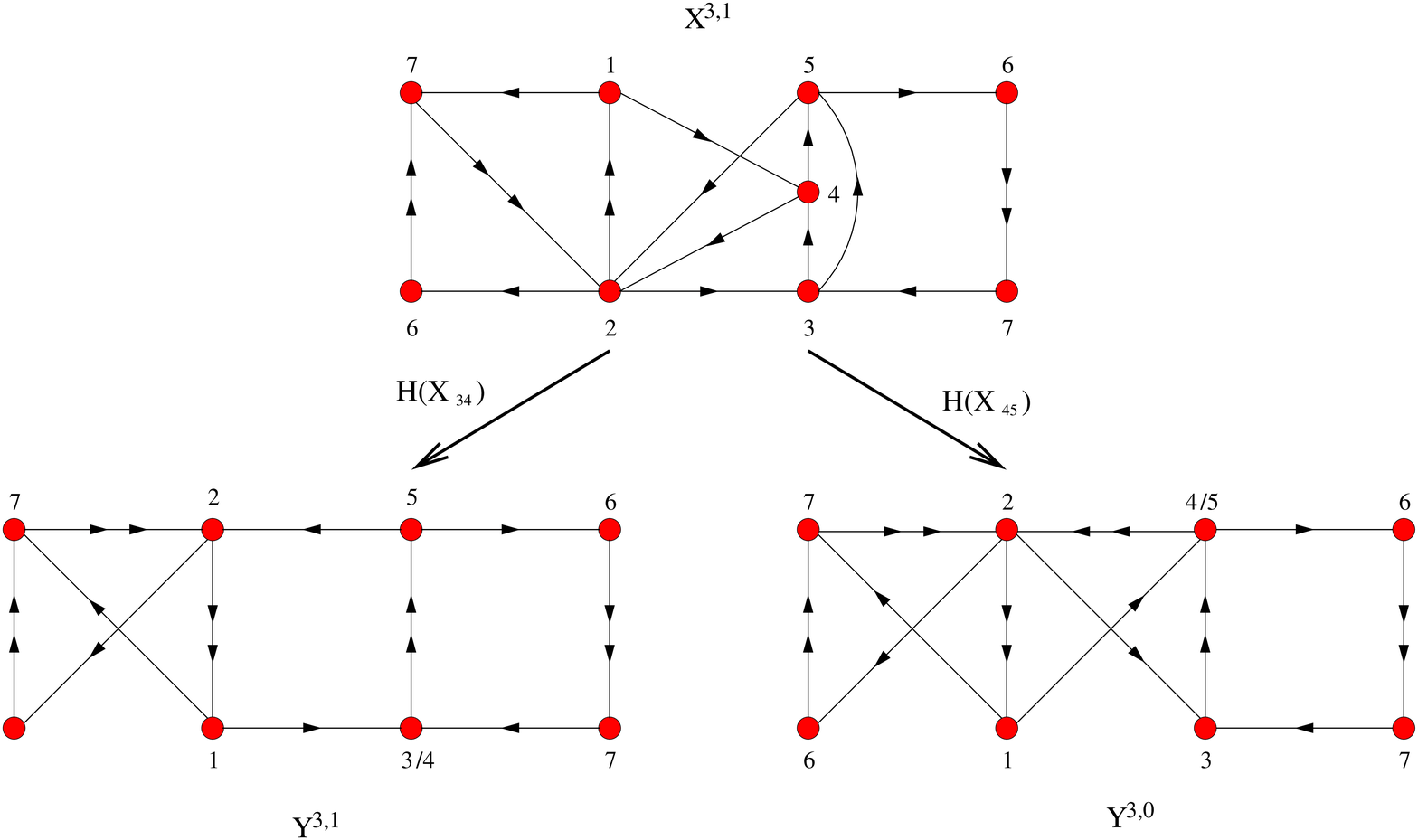}}
 \caption{One of the two toric phases of $X^{3,1}$ with 17 fields.  The
notation $H(X_{ij})$ means giving a vev to the scalar component of
$X_{ij}$.}
 \label{17fields}
\end{figure}

The phases of $X^{3,1}$ we have described so far have either 15 or 19
fields.  When we Higgs  them down to $Y^{3,1}$ only the field that gets a
vev disappears and
we get the phases with with 14 and 18 fields respectively. Higgsing to
$Y^{3,0}$ gives mass to two fields which are integrated out and removed
from the massless spectrum
together with the field that gets the expectation value, producing the
toric phases with 12 and 16 fields.  The difference in the
number of fields simply comes from the difference in the distribution of
impurities
in  $Y^{3,1}$. Whenever two single impurities fuse into a double impurity
in $Y^{p,q}$, the number of fields goes up by four. In addition to these,
there are
also ``intermediate" toric phases of $X^{3,1}$ that have 17 fields.  These
blow down to the phase of $Y^{3,1}$ that  has 14 fields and to the phase
of  $Y^{3,0}$
with 16 fields.  The details of the Higgsing are now reversed: two
additional fields are integrated out when going to $Y^{3,1}$, but only the
field with a vev disappears when
we blow down to $Y^{3,0}$.  For $X^{3,1}$ there are precisely two such
phases, produced by dualizing nodes 1 or 3.   In \fref{17fields} we show
the
first of these two, and the phases of $Y^{p,q}$ to which it blows down.

\begin{figure}[ht]
 \epsfxsize = 6cm
 \centerline{\epsfbox{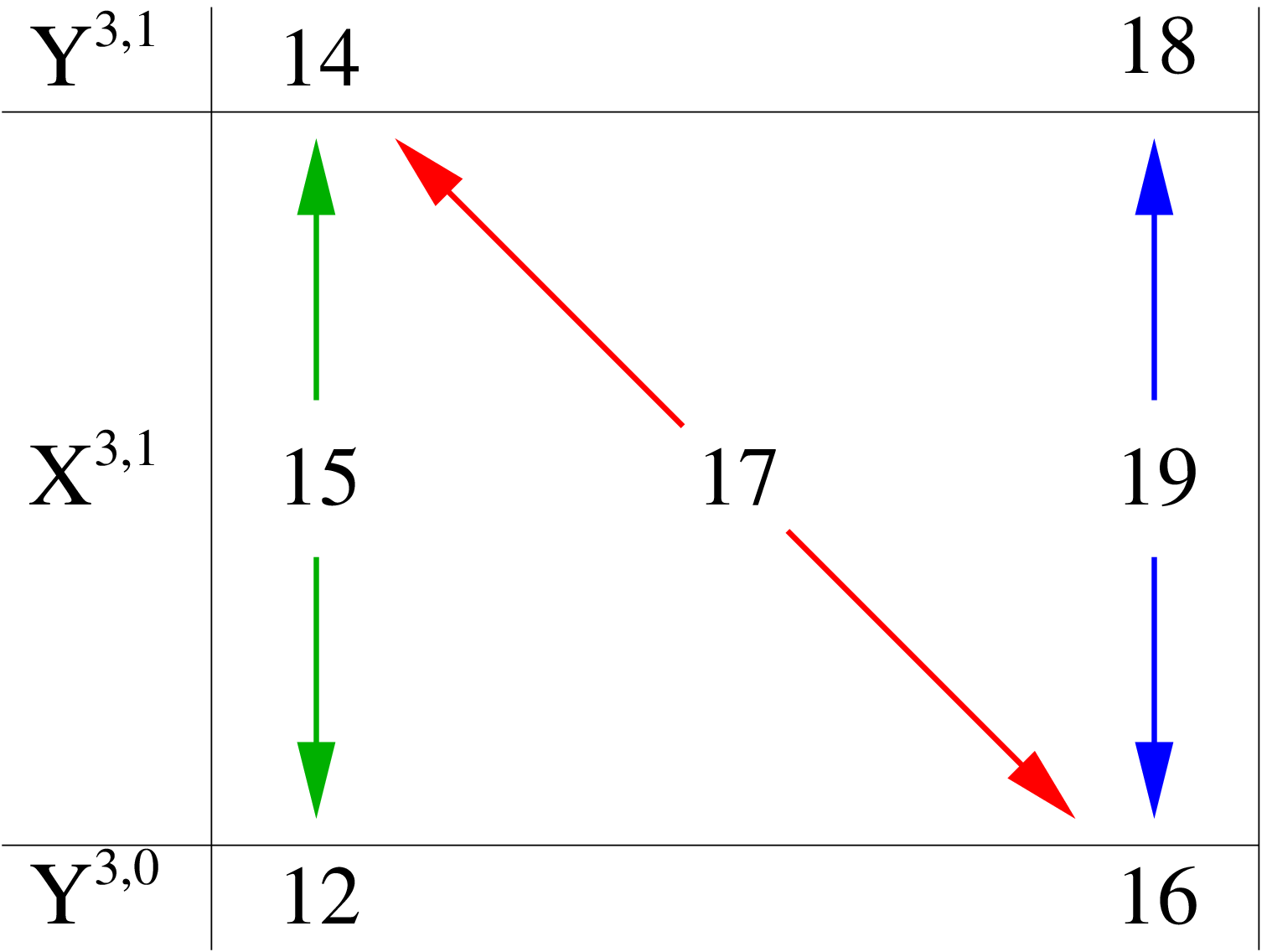}}
 \caption{The number of fields for the toric phases of $X^{31}, Y^{31}, Y^{30}$ 
 and the possible higgsings (shown with arrows). Note that several different toric 
 phases can have the same number of fields.  }
 \label{higgsings}
\end{figure}

More generally, we expect the toric phases of all $X^{p,q}$ to fall in
this pattern. Increasing $p$ does not change the story, because
the blow up is localized on the quiver graph. The $Y^{p,q}$ theories have
toric phases with $4p+2q+4n_2$ fields
where $n_2=0,1,\ldots \left[\frac{p-q}{2}\right]$ is the  number of double
impurities.  For each of these models there will be toric phases of
$X^{p,q}$
with one additional field that Higgs down to $Y^{p,q}$ and $Y^{p,q-1}$
like the examples in Figures \ref{doubleimpurity},\ref{dualize4}.  On top of
these we have
the ``intermediate'' phases with $4p+2q+4n_2+3$ fields, where $n_2=0,1,\ldots
\left[\frac{p-q}{2}\right]-1$. These also blow down to $Y^{p,q}$,
$Y^{p,q-1}$ in the way
described for the example in \fref{17fields}. This is summarized in 
in \fref{higgsings} for the case of $X^{3,1}$. It would be interesting to
have a more general understanding of the Seiberg dual phases of $X^{p,q}$,
including
non-toric ones, since it may be easier to extract information about the
infrared dynamics from some of these models than from others.

The toric phases of $Y^{p,q}$ can  be easily enumerated.  The problem is the same 
as counting the ways of coloring the $p$ vertices of a $p$-gon using three different 
colors (corresponding to a single impurity, a double impurity, or no impurity) modulo 
the action of the dihedral group $D_p$. This is
a standard application of P\'{o}lya's enumeration theorem \cite{Polya}. The cycle index of the 
dihedral group is 
\bea
\mathcal{Z}(D_p)&=&\frac{1}{2}\mathcal{Z}(Z_p) + \frac{1}{2}x_1 x_2^{(p-1)/2}\;\;\;\;\;\;\;\;\;\;\;\;\;\;\;\;\;\;(p \;\;\mbox{odd})\nn \\
\mathcal{Z}(D_p)&=&\frac{1}{2}\mathcal{Z}(Z_p) + \frac{1}{4}(x_2^{p/2}+x_1^2 x_2^{(p-2)/2})\;\;\;\;\;(p\;\; \mbox{even})
\label{cycledp}
\eea
where $\mathcal{Z}(Z_p)$ is the cycle index of the cyclic group of order $p$ :
\bea
\mathcal{Z}(Z_p)=\frac{1}{p}\sum_{n|p}\phi(n)x_n^{p/n}
\eea
and $\phi(n)$ is Euler's totient function.  To implement the condition $n_1+2n_2=p-q$ we assign 
weight $1$ for no impurity, $\lambda$ to a single impurity and $\lambda^2$ to a double impurity.  
Then $x_n=1+\lambda^n+\lambda^{2n}$, and after plugging this into (\ref{cycledp}) the number of toric phases is 
given by the coefficient of $\lambda^{p-q}$ in the resulting polynomial.  Because of the way the $X^{p,q}$ 
theories Higgs to $Y^{p,q}$ and $Y^{p,q-1}$, we expect the number of toric phases of $X^{p,q}$ to be essentially
determined by (and in fact greater than) the number of toric phases of  $Y^{p,q}$, 
but we shall leave a detailed counting to future work.

Finally, we note that when performing Seiberg duality one must alter
the superpotential as well as the quiver. This is done
by including cubic interactions of the form $Mq\tilde q$, where
$q$ and $\tilde q$ are the dual quarks and $M$ is a composite field
in the original theory which maps to a singlet in the dual. One must be
careful when applying this procedure to bifundamentals, since here the field
$M$ is only a singlet under the dualized group but still transforms as a bifundamental
in the dual theory. This procedure works as expected for the $X^{p,q}$ theories
and produces the necessary $Y^{p,q}$ superpotentials upon giving a vev to
the appropriate bifundamentals. For more details, we refer the reader to
\cite{Beasley:2001zp,Feng:2001bn}.

%===========================================================
\section{R-charges}
%===========================================================

An important test of the AdS/CFT correspondence is that the dimensions
of operators in the field theory match with the volumes of the different 
supersymmetric (calibrated) submanifolds
D3-branes can wrap in the geometry \cite{Witten:1998xy,Berenstein:2002ke}. 
These dimensions, or equivalently R-charges, 
can be figured out in
the gauge theory via $a$-maximization, and in the geometry via a volume calculation.
These charges have also been computed via techniques from algebraic geometry 
\cite{iwbaryon,Herzog:2003zc,Herzog:2003dj,Herzog:2003wt}.
It is also possible to derive some of the R-charges purely from the toric data \cite{djy}
\footnote{We thank James Sparks and Dario Martelli for alerting us to this
work, and also for their comments, especially on this section of this paper.}.

Let's briefly review the situation for $dP_2$, which will reveal an interesting
aspect of the $X^{p,q}$ theories. 
The quiver for $dP_2$ is given in \fref{quiver_dp2}. The R-charges for this 
theory are known \cite{Bertolini:2004xf}, and given in Table \ref{charges}. 
Since baryonic operators in the SCFT correspond to D3-branes wrapping 3-cycles,
we may associate to each bifundamental a 3-manifold in the dual geometry. 
Since $dP_2$ has two 3-cycles, we expect two
baryonic $U(1)$ global symmetries. 
Similarly, we expect $dP_2$ to have 
a $U(1) \times U(1)$ isometry, which
should translate to an additional $U(1)^2$ flavor symmetry.
The baryonic $U(1)$'s do not
mix with the R-symmetry, 
as discussed in \cite{Intriligator:2003jj,iwbaryon}. Therefore 
there should be a two-dimensional space
of $U(1)$ symmetries which can potentially mix with
the R-symmetry. Indeed, we note that we may assign a 2-dimensional basis of R-charges
to the fields in $dP_2$ as given in Table \ref{charges}. 
One can see a similar agreement for the $Y^{p,q}$ theories:
With one non-baryonic $U(1)$ symmetry, one can
reduce the R-charges to a one-dimensional basis. This
is what allowed the authors of \cite{BFHMS} to compute the R-charges
of these theories.

Equivalently, we can rephrase the above discussion as follows:
Since the baryonic symmetries do not mix with the R-charges,
it should be possible to do $a$-maximization over a two-dimensional
space of parameters
for $dP_2$ and over a one-dimensional space for $Y^{p,q}$. 
Naively, one might think from the field theory that $dP_2$ would
require $a$-maximization over four parameters; this is
what comes out of solving the linear equations
given by the constraints of anomaly freedom at each node,
and that the superpotential terms all have R-charge equal to two.
However, we know that this is not the whole story; if one were
able to pick the two flat directions properly, $a$-maximization
could be done over only two parameters. This is
easy to do once we know the right R-charges.
From these, it is possible to work backwards and assign
a good two-dimensional basis of charges as in Table \ref{charges}.
Doing $a$-maxmization over these two-dimensional basis, treating these
charges as free parameters, yields the correct result.

\begin{table}[!h]
\begin{center}
$$\begin{array}{|c|c|c|}  \hline
 \mathrm{Field}  & R -\mathrm{charge} & \mathrm{Linear\,\, Combination} \\ \hline\hline
X_{52} &  \frac{3}{16}(19-3\sqrt{33})     &\Sigma_1  \\\hline
		X_{14} &  \frac{3}{16}(19-3\sqrt{33})     &\Sigma_1  \\\hline
	X_{53} &  \frac{1}{4}(9-\sqrt{33})     & \Sigma_2  \\\hline
		X_{34} &  \frac{1}{4}(9-\sqrt{33})     &\Sigma_2  \\\hline   
 X_{42} &  \frac{1}{16}(-21+5\sqrt{33})     
& \Sigma_2-\Sigma_1  \\\hline
     X_{51} &  \frac{1}{16}(-21+5\sqrt{33})     
& \Sigma_2-\Sigma_1  \\\hline
      X_{23}^1 &  \frac{1}{16}(-21+5\sqrt{33})     
& \Sigma_2-\Sigma_1  \\\hline
       X_{31}^1 &  \frac{1}{16}(-21+5\sqrt{33})     & 
\Sigma_2-\Sigma_1  \\\hline
        X_{23}^2 &  \frac{1}{16}(17-\sqrt{33})     & \Sigma_2-\frac{1}{3}\Sigma_1  \\\hline
          X_{31}^2 &  \frac{1}{16}(17-\sqrt{33})     &\Sigma_2-\frac{1}{3}\Sigma_1  \\\hline
	
	    X_{45} &  \frac{1}{2}(-5+\sqrt{33})     & \Sigma_2-\frac{4}{3}\Sigma_1  \\ \hline
\end{array}$$
\caption{R-charges for $dP_2$. All R-charges can be expressed as a linear combination of
two basis charges.}
\label{charges}
\end{center}
\end{table}

One would naively think that just as the $dP_1$ results extended to general $Y^{p,q}$, 
we could extend the $dP_2$ results to general $X^{p,q}$. Since
the number of baryonic $U(1)$'s is always the number of external legs of
the toric diagram
minus three, there should always be two $U(1)_B$ symmetries
in the $X^{p,q}$ theories that do not mix with the R-symmetry. 
Since we expect there to be naively four free parameters for
the $X^{p,q}$ theories in general (we have verified this for e.g. $X^{3,1}$
although not for general $X^{p,q}$),
there should be a remaining two-dimensional basis of R-charges.

Unfortunately, however,
we have not been able to reduce the
R-charges of these theories 
to a two-dimensional basis.  One difficulty is that since $a$-maximization for the
$X^{p,q}$ theories must be done over
four parameters (at least initially, since it
is difficult to pick the flat directions in practice), it is very difficult to obtain exact numbers for 
$p > 2$. We have, however, numerically computed the R-charges for 
$X^{3,q}$. Since the numbers are not particularly illuminating, we do not record
the R-charges here. We do note that it appears {\bf not} to be true
that there is a two-dimensional basis of R-charges for general $X^{p,q}$ theories.
One can easily check if, given three R-charges, there is any integer linear combination
of them that equals another integer. To the precision allowed by Mathematica, we have
not found any such linear combination of R-charges for $X^{3,q}$ for any $1 \leq q \leq 3.$
Thus, it appears that our naive guess that there are only two $U(1)$ flavor symmetries  
in the Einstein-Sasaki
manifold
is incorrect.
We note that the
$X^{p,p}$ theory is special in that the quiver and superpotential have a 
$\IZ_2$ symmetry
which gives a nontrivial constraint on the R-charges; in these
cases we may reduce the number of independent R-charges to three. Thus,
we have a puzzle: Since the baryonic $U(1)$'s should not mix with
the R-symmetry, there should
be relations between the R-charges. Our inability to find such relations may
be a consequence of our numerical computation; since we cannot
find the {\bf exact} R-charges for the $X^{p,q}$, we can only
check that things sum to zero up to a given numerical precision. However,
it is possible that there is a deeper issue here as well. 

Finally, we note that there are some $X^{p,q}$ where we can actually find the
R-charges exactly. We have done the calculation for $X^{2,2}$ and $X^{3,3}$, and found
that in these cases the R-charges are not quadratic irrational, but instead the roots
of {\bf quartic} polynomials. These are the first examples of theories whose
R-charges are not quadratic irrationals. This is not in contradiction to 
the prediction from $a$-maximization, since the charges are still
algebraic numbers. Since $a$ is a cubic function over many variables, derivatives
of $a$ will be quadratic functions over many variables. In general the solutions
of these equations will 
not be quadratic irrational, although they have been for every case studied so far.
Although we do not record the exact values of all the R-charges for $X^{2,2}$ and $X^{3,3}$ 
here, we do note for the sake of reference two R-charges. For $X^{3,3}$ (shown in 
\fref{x3q}), the R-charge of the bifundamental $X_{56}$ is given by a 
root of the polynomial $27x^4- 198x^3 - 180x^2+ 650x-250$. For $X^{2,2}$
(also known as Pseudo-del Pezzo 2), the R-charge of one of the fields
($X_{53}$ in Figure 6 of \cite{Feng:2004uq}) is given by a root of 
$9x^4- 78x^3+ 112x^2+ 16x+32$.

%===========================================================
\section{Acknowledgements}
%===========================================================

We would like to thank Alan Dunn, Sebastian Franco, Dan Freedman, Barak Kol, Dario
Martelli, and James Sparks for awesome discussions. BW would like
to thank Nick Jones and his amusing accent for last-minute TeX help, and acknowledge the
hospitality of Brook Williams's apartment, where the final
stages of this work were completed. 
This work was supported in part by the CTP and LNS of
MIT and the U.S. Department of Energy under cooperative research agreement $\#$ DE--FC02--
94ER40818, National Science Foundation Grant PHY-00-96515,
and by the BSF American--Israeli Bi--National Science Foundation. A. H. is also
indebted to a DOE OJI Award.

%%%%%%%%%%%%%%%%%%%%%%%%%%%%%%%%%%%%%%%%%%%%%%%%%%%%%%%%%%%%%%%%%%%%%%%%%%%%%%%%%%%%%%%%%%%%%%%%%%%%%%%%%%%%%%%%%%%%%%%%%%%%%%%%%%%%%%%%%%%%%%%%%%%%%%%

\bibliographystyle{JHEP}

\typeout{LaTeX Warning: Label(s) may have changed. Rerun}
\end{document}